\documentclass[usenatbib]{mn2e}

\usepackage{epsfig}
\usepackage{graphics}
\usepackage{afterpage}

\newlength{\colwidth}
\newcommand{\cm}{{\rm cm}}

\newcommand{\kms}{{\rm km}\,{\rm s}^{-1}}
\newcommand{\K}{{\rm K}}
\newcommand{\pc}{{\rm pc}}
\newcommand{\hkpc}{h^{-1}\,{\rm kpc}}

\newcommand{\hMpc}{h^{-1}\,{\rm Mpc}}
\newcommand{\erg}{{\rm erg}}
\newcommand{\hMsun}{{h^{-1}\,{\rm M}_\odot}}
\newcommand{\Msun}{{{\rm M}_\odot}}

\newcommand{\nH}{{{\rm n}_{\rm H}}}
\newcommand{\nHs}{{{\rm n}_{\rm H}^*}}

\newcommand{\yr}{{\rm yr}}

\newcommand{\Msolyrkpcsq}{{\rm M}_\odot\,{\rm yr}^{-1}\,{\rm kpc}^{-2}}
\newcommand{\ion}[2]{\hbox{#1\,{\sc #2}}}

\newcommand{\CIV}{\ion{C}{IV}}

\newcommand{\NaI}{\ion{Na}{I}}

\title[The physics driving the star formation history]{The physics
  driving the cosmic star formation history} 
\author[J. Schaye et al.]{%
Joop~Schaye,$^1$\thanks{E-mail: schaye@strw.leidenuniv.nl} 
Claudio~Dalla~Vecchia,$^1$ C.~M.~Booth,$^1$ 
Robert~P.~C.~Wiersma,$^1$ \newauthor
Tom~Theuns,$^{2,3}$ 
Marcel~R.~Haas,$^1$ Serena~Bertone,$^4$ Alan~R.~Duffy,$^{1,5}$ 
\newauthor
I.~G.~McCarthy,$^6$
and Freeke~van~de~Voort$^1$\\  
\\
$^{1}$ Leiden Observatory, Leiden University, P.O. Box 9513, 2300 RA Leiden,
  the Netherlands\\
$^2$ Institute for Computational Cosmology, Department of Physics,
University of Durham, South Road, Durham, DH1 3LE, UK\\
$^3$ Department of Physics, University of Antwerp, Groenenborgerlaan
171, B-2020 Antwerpen, Belgium\\
$^4$ Santa Cruz Institute for Particle Physics, University of
California at Santa Cruz, 1156 High Street, Santa Cruz, CA 95064, USA\\
$^5$ Jodrell Bank Centre for Astrophysics, Alan Turing Building, The
University of Manchester, Manchester M13 9PL\\ 
$^6$ Kavli Institute for Cosmology, University of Cambridge, Madingley
Road, Cambridge CB3 0HA
}

\begin{document}

\pagerange{\pageref{firstpage}--\pageref{lastpage}} \pubyear{2007}

\maketitle

\label{firstpage}

\begin{abstract}
We investigate the physics driving the cosmic star formation (SF)
history using the more than fifty large, cosmological, hydrodynamical
simulations that together comprise the OverWhelmingly Large
Simulations (OWLS) project. We systematically vary the parameters of
the model to determine which physical processes are dominant and which
aspects of the model are robust. Generically, we find that SF is
limited by the build-up of dark matter haloes at high redshift,
reaches a broad maximum at intermediate redshift, then decreases as it
is quenched by lower cooling rates in hotter and lower density gas,
gas exhaustion, and self-regulated feedback from stars and black
holes.  The higher redshift SF is therefore mostly determined by the
cosmological parameters and to a lesser extent by photo-heating from
reionization.  The location and height of the peak in the SF history,
and the steepness of the decline towards the present, depend on the
physics and implementation of stellar and black hole feedback.  Mass
loss from intermediate-mass stars and metal-line cooling both boost
the SF rate at late times.  Galaxies form stars in a self-regulated
fashion at a rate controlled by the balance between, on the one hand,
feedback from massive stars and black holes and, on the other hand,
gas cooling and accretion. Paradoxically, the SF rate is highly
insensitive to the assumed SF law. This can be understood in terms of
self-regulation: if the SF efficiency is changed, then galaxies adjust
their gas fractions so as to achieve the same rate of production of
massive stars. Self-regulated feedback from accreting black holes is
required to match the steep decline in the observed SF rate below
redshift two, although more extreme feedback from SF, for example in
the form of a top-heavy initial stellar mass function at high gas
pressures, can help.
\end{abstract}


\section{Introduction}

The cosmic history of star formation (SF) is one of the most fundamental
observables of our Universe. Measuring the global star formation rate
(SFR) density as a function of redshift has therefore long been one of
the primary goals of observational astronomy
\citep[e.g.][]{Lilly1996SFH,Madau1996SFH,Steidel1999SFH,Ouchi2004SFH,Schiminovich2005SFH,Arnouts2005SFH,Hopkins2006SFH,Bouwens2007UVLFs}. 

Modeling
the cosmic star formation history (SFH) is not an easy task because it
depends on a complex interplay of physical processes and because a
large range of halo masses contributes. To predict the SFH
within the context of the cold dark matter cosmology, one must first
get the dark matter halo mass function right. These days this
is the easier part, as the cosmological parameters are relatively well
constrained. Then one must model the rate at which gas accretes,
cools, collapses, and turns into stars. Even if one does not attempt
to model the cold, interstellar gas phase and uses empirical SF
laws to estimate the rate at which gas is converted
into stars on kpc scales, there are a 
host of feedback processes that need to be taken into account. Stars
produce radiation which can 
heat gas, exert radiation pressure, and change its ionization balance
and hence its cooling rate. Massive stars explode as supernovae (SNe)
which can drive both small-scale turbulence and large-scale
outflows. Stars also produce heavy 
elements and dust which change the rate at which gas cools. Accretion
onto supermassive black holes (BHs) in the centers of 
galaxies also results in radiative, thermal and mechanical
feedback. Finally, magnetic fields and cosmic rays may be important. 

Despite this complexity, many authors have used semi-analytic 
models
\citep[e.g.][]{White1991GF,Hernquist2003SFH,Bower2006SA,Croton2006SA,Somerville2008SA,Fontanot2009SA}
or hydrodynamical simulations 
\citep[e.g.][]{Nagamine2000SFH,Pearce2001Galform,Ascasibar2002SFH,Murali2002SFH,Springel2003SFH,SommerLarsen2003Cosmosim,Keres2005Gasaccr,Nagamine2006SFH,Oppenheimer2008Wmom,DiMatteo2008AGNCosm,Choi2009Metalcooling,Crain2009Gimic,Booth2009AGN}
to study the SFH of the Universe. 

Modern semi-analytic models use
cosmological dark matter simulations to generate halo mass functions
and merger trees. These are then combined with simple prescriptions
for the baryonic physics, which include a large number of free
parameters, to predict galaxy SFRs. The parameters of the models are 
tuned to match particular sets of observations, after which the model
is used to make predictions for other properties. While this approach
has been very productive, it can be difficult to understand what
physics is driving the results and there is a danger that model
predictions may be correct even when the underlying prescriptions do not
reflect the real world (because they were tuned to match particular
observations). Furthermore, current semi-analytic models are only of
limited use for the study of the gas around and in between galaxies.  

While hydrodynamical simulations
attempt to model much more physics from first principles than
semi-analytic models, they do make extensive use of
subgrid prescriptions for the physics that they cannot
resolve directly. Some of these prescriptions are physically motivated
and relatively well understood. One example is the radiative cooling
rate which is determined by atomic physics. However, even here there are
significant uncertainties such as the effects of
photo-ionization, non-equilibrium, dust, and relative abundance
variations. Other subgrid prescriptions, such as the stellar initial mass
function (IMF), are empirically
motivated. Most subgrid models contain, however, a mixture of physical
and empirical ingredients. For example,
prescriptions for SF, thermal/kinetic feedback from
stars and active galactic nuclei (AGN), and for the growth of BHs
typically use physically 
motivated frameworks whose parameters are calibrated using
observations. 

When it comes to predictions for SFRs in galaxies, the subgrid models
used in cosmological hydro simulations play a critical role. Differences in
the subgrid prescriptions used by different authors are likely to be
much more important than differences in the codes used to model gravity and
hydrodynamics. The crucial role played by subgrid models implies that 
hydrodynamical simulations have some of the same weaknesses as semi-analytic
models. On the other hand, direct simulation enables one to probe
much deeper, e.g.\ by following gas flows in three dimensions, or by
studying the interactions of galaxies with their environment and the
intergalactic medium (IGM).  

While simulations offer the tantalizing possibility of laboratory-like
control, the huge dynamic range required makes it computationally very
expensive to explore parameter space. The OverWhelmingly Large
Simulations (OWLS) project, which was made possible by the
temporary availability of the supercomputer that serves as the
correlator for the LOFAR telescope, aims to use the potential of
simulations to gain insight into the physics that determines the
formation of galaxies and the evolution of the IGM. The OWLS project
consists of a large suite of cosmological, smoothed particle
hydrodynamics (SPH) simulations with
varying box sizes and resolutions. Each production simulation
uses $2\times 512^3$ particles, which places them
among the largest dissipative simulation ever performed. The real power of
the project stems, however, not from the size of the simulations, but
from the fact that they are repeated many (more than 50) 
times, each time varying a subgrid prescription, most of which were
newly developed for this project. Although we have not
attempted to fine-tune the subgrid parameters 
to match particular data sets, we do hope that our investigations
will help trigger future work in directions that will improve
agreement with the observations. 

While the OWLS project aims to study a variety of problems in
astrophysical cosmology, we will limit ourselves here to the cosmic
SFH. To further limit the scope of the paper, we will postpone a
discussion of the SFR as a function of halo mass to another paper
(Haas et al., in preparation). We note, however, that
this latter function is in some ways physically more interesting than
the SFH. This is because the cosmic SFR can be thought of as a
convolution of the halo mass function, which depends only on cosmology
and redshift, and the SFR as a function of halo mass and redshift,
which depends on poorly understood astrophysics.

\begin{table*} 
\begin{center}
\caption{List of simulations run assuming the reference model.  From
  left-to-right the columns show: simulation identifier; comoving box size;
  number of dark matter particles (there are equally 
  many baryonic particles); baryonic particle mass; dark matter
  particle mass; comoving (Plummer-equivalent) gravitational
  softening; maximum physical softening; final redshift. } 
\label{tbl:ref_sims}
\begin{tabular}{lrrrlrrl}
\hline
Simulation & $L$ & $N$ & $m_{\rm b}$ & $m_{\rm dm}$ &
$\epsilon_{\rm com}$ & $\epsilon_{\rm prop}$ & $z_{\rm end}$\\  
& $(\hMpc)$ & & $(\hMsun)$ & $(\hMsun)$ & $(\hkpc)$ &
$(\hkpc)$ & \\
\hline 
\emph{REF\_L006N128} &   6.25 & $128^3$ & $1.4 \times 10^6$
& $ 6.3 \times 10^6$ & 1.95 & 0.50 & 2\\
\emph{REF\_L012N256} &  12.50 & $256^3$ & $1.4 \times 10^6$
& $ 6.3 \times 10^6$ & 1.95 & 0.50 & 2 \\
\emph{REF\_L012N512} &  12.50 & $512^3$ & $1.7 \times 10^5$
& $ 7.9 \times 10^5$ & 0.98 & 0.25 & 2 \\
\emph{REF\_L025N128} &  25.00 & $128^3$ & $8.7 \times 10^7$
& $ 4.1 \times 10^8$ & 7.81 & 2.00 & 0 \\
\emph{REF\_L025N256} &  25.00 & $256^3$ & $1.1 \times 10^7$
& $ 5.1 \times 10^7$ & 3.91 & 1.00 & 2 \\
\emph{REF\_L025N512} &  25.00 & $512^3$ & $1.4 \times 10^6$
& $ 6.3 \times 10^6$ & 1.95 & 0.50 & 1.45 \\
\emph{REF\_L050N256} &  50.00 & $256^3$ & $8.7 \times 10^7$
& $ 4.1 \times 10^8$ & 7.81 & 2.00 & 0 \\
\emph{REF\_L050N512} &  50.00 & $512^3$ & $1.1 \times 10^7$
& $ 5.1 \times 10^7$ & 3.91 & 1.00 & 0 \\
\emph{REF\_L100N128} & 100.00 & $128^3$ & $5.5 \times 10^9$
& $ 2.6 \times 10^{10}$ & 31.25 & 8.00 & 0 \\
\emph{REF\_L100N256} & 100.00 & $256^3$ & $6.9 \times 10^8$
& $ 3.2 \times 10^9$ & 15.62 & 4.00 & 0 \\
\emph{REF\_L100N512} & 100.00 & $512^3$ & $8.7 \times 10^7$
& $ 4.1 \times 10^8$ & 7.81 & 2.00 & 0 \\
\hline
\end{tabular}
\end{center}
\end{table*}

We will also not compare to observations of the build up of the cosmic
stellar mass density. The cumulative SFR
must of course equal the stellar mass (after taking stellar mass loss
into account). While this will be true by
construction for ab initio models, it is not necessarily the case
for observationally inferred quantities. This is because observational
probes of SF measure only the rate 
of formation of massive stars, since those dominate the
electromagnetic output of young stellar populations. To obtain the
total SFR, it is necessary to extrapolate to low masses using an
assumed IMF. Since the IMF is uncertain and may not be universal,
measurements of the total stellar mass are of great interest. However,
it should be noted that massive stars are the
ones that matter from a cosmological perspective because they dominate
the chemical, radiative and mechanical feedback processes that
regulate the formation of stars and galaxies. 
As the objective of this paper is to explore the 
effects of varying physical prescriptions rather than to fit the
observations, we will for simplicity limit ourselves to comparisons
with measurements of the evolution of the SFR density. We emphasize,
however, that it should be kept in mind that these are often
inconsistent with observations of the stellar mass density
(e.g.\ \citealt{Wilkins2008SFH,Cowie2008SFH}, but see also
\citealt{Reddy2008SFH,Reddy2009SFH}) and subject to large systematic
uncertainties \citep[e.g.][]{Conroy2009uncertainties}.   

This paper is organized as follows. 
In section~\ref{sec:simulations} we introduce the numerical techniques
that are common to all OWLS runs, while sections~\ref{sec:reference}
and \ref{sec:variations} describe the physics modules employed in the
reference model and the other OWLS runs, respectively. As this
paper also serves to introduce the OWLS project, we will
discuss the ingredients of the different simulations in some detail. The SF
histories predicted by the models are presented and discussed in the
sections that describe them. 
Finally, we discuss and summarise our main findings in
section~\ref{sec:summary}. 

\section{Simulations}
\label{sec:simulations}

In this section we describe the numerical techniques that are common
to all simulations that make up the
OWLS project. The modifications and additions to the subgrid modules are
discussed in section~\ref{sec:reference} for our reference model and
in section~\ref{sec:variations} for the other runs.

The simulations were performed with a significantly
extended version of the $N$-Body Tree-PM, SPH code \textsc{gadget3} 
\citep[last described in][]{Springel2005Gadget2}, a Lagrangian code
used to calculate gravitational and hydrodynamical forces on a system of
particles. 
Most simulations were run in periodic boxes of size $L=25$ and
100 comoving $\hMpc$ and each of the production runs
uses $512^3$ dark matter and equally many baryonic particles
(representing either collisionless star or collisional gas
particles). The particle masses in the $2\times512^3$ particle
25~$\hMpc$ (100~$\hMpc$) box are $6.34 \times 10^{6}\,\hMsun$
($4.06\times 
10^8\,\hMsun$) for dark matter and $1.35 \times 10^{6}\,\hMsun$
($8.66\times10^7\,\hMsun$) for baryons.  Note, however, that baryonic
particle masses change during the course of the simulation due to mass
transfer from star to gas particles.  The 25~$\hMpc$
simulation volumes were run as far as redshift 2 and the 100~$\hMpc$
volumes were run to redshift 0. Comoving gravitational softenings were
set to $1/25$ of the initial mean inter-particle spacing but were
limited to a maximum physical scale of 0.5~kpc$/h$ (2~kpc$/h$) for the
25~$\hMpc$ (100~$\hMpc$) simulations. The switch from a fixed comoving
to a fixed proper softening happens at $z=2.91$ in all
simulations. We used $N_{\rm ngb} = 48$ neighbors for the SPH interpolation.   

In order to assess the effects
of the finite resolution and box size on our results, we have run
a suite of cosmological simulations, all using the same physical
model, using different box sizes (ranging from 
6.25~$\hMpc$
to 100~$\hMpc$) and particle numbers (ranging from $128^3$ to
$512^3$).  The particle masses and gravitational softenings for each
of these 
simulations are listed in Table~\ref{tbl:ref_sims}. 

The initial conditions were generated with \textsc{cmbfast} (version
4.1; \citealt{Seljak1996CMBFAST}) and evolved to redshift $z=127$,
where the simulations were started, 
using the \citet{Zeldovich1970Approx} approximation from an initial
glass-like state \citep{White1996glass}.

Before discussing each of the
variations made to the subgrid models in section~\ref{sec:variations},
we first turn out attention to the physics included in the reference model.

\section{The reference model}
\label{sec:reference}

\subsection{Description of the model}
\label{sec:ref_descr}

When simulations lack the required numerical
resolution or the physics to accurately model a physical process, we
must resort to subgrid models.  In this section we describe the
\lq reference\rq\, physical model (\emph{REF}) that is used as a base
from which 
all further investigations of the behaviour of the simulations can be
launched. As discussed in section~\ref{sec:variations}, we will do this by
varying one physical process at a time and comparing the resulting
SFH to the one predicted by the reference model. 

We emphasize that the \emph{REF} model should not be viewed as our
``best'' model. As its name implies, it functions as a reference point
for our systematic variation of the parameters. In fact, we will show in
future papers that some variations yield much better agreement with
particular types of observations. For example, the inclusion of AGN
feedback dramatically improves the agreement with observations of
groups of galaxies at redshift zero (McCarthy et al., in preparation).  

\begin{figure*}
\resizebox{\textwidth}{!}{\includegraphics{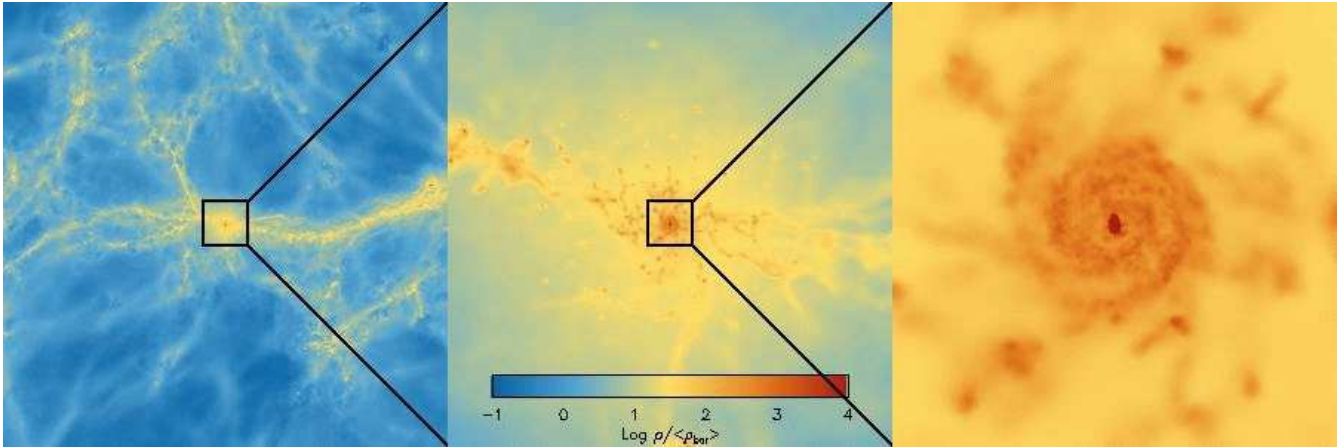}}
\caption{Zoom into a $M_{200} = 10^{12.2}\,\Msun$ halo at $z=2$ in the
  \emph{REF\_L025N512} simulation. From left-to-right, the images are
  10, 1, and 0.1~$\hMpc$ on a side. All slices are 1~$\hMpc$
  thick. Note that the first image shows only a fraction of the total
  simulation volume, which is cubic and 25~$\hMpc$ on a side. The
  color coding shows the projected gas density, $\log_{10}
  \rho/\left <\rho_{\rm b}\right >$, and the color scale ranges from 
  -1 to 4 (which is lower than the true maximum of the image). The
  coordinate axes were rotated to show the galaxy face-on. This
  halo is the 10th most massive in the simulation. About half of the
  haloes in this mass range host extended disk galaxies, while the
  other half have highly disturbed morphologies due to ongoing
  mergers.}
\label{fig:zoomL025}
\end{figure*}

\begin{figure*}
\resizebox{\textwidth}{!}{\includegraphics{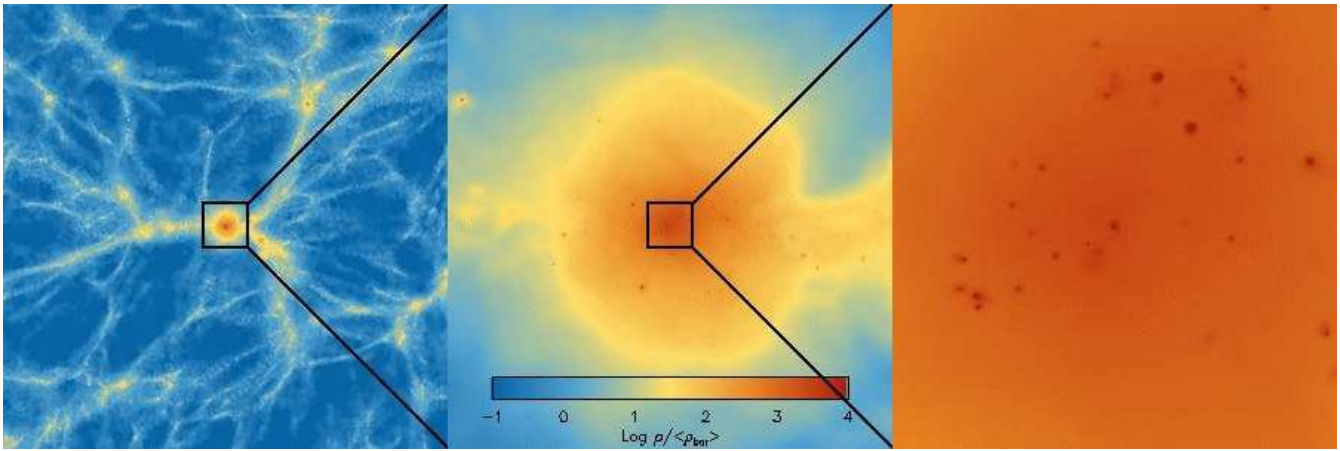}}
\caption{Zoom into a $M_{200} = 10^{14.2}\,\Msun$ halo at $z=0$ in the
  \emph{REF\_L100N512} simulation. From left-to-right, the images are
  40, 4, and 0.4~$\hMpc$ on a side. All slices are 1~$\hMpc$
  thick. Note that the first image shows only a fraction of the total
  simulation volume, which is cubic and 100~$\hMpc$ on a side. The
  color coding shows the projected gas density, $\log_{10}
  \rho/\left <\rho_{\rm b}\right >$, and the color scale ranges from 
  -1 to 4 (which is lower than the true maximum of the image). This
  halo is the 10th most massive in the simulation.}
\label{fig:zoomL100}
\end{figure*}

To illustrate the dynamic
range in the simulations, Figs.~\ref{fig:zoomL025} and
\ref{fig:zoomL100} show two factors of ten zooms into the tenth most 
massive haloes in, respectively, the \emph{REF\_L025N512} run at $z=2$
and model \emph{REF\_L100N512} at $z=0$. Note, however, that the left 
panels still show only a small fraction of the simulation volume.


\subsubsection{Cosmology}

We assume the values for the cosmological parameters derived from the
Wilkinson Microwave Anisotropy Probe (WMAP) 3-year results
\citep{Spergel2007Wmap3},  
$\{\Omega_{\rm m}, \Omega_{\rm b}, \Omega_\Lambda, \sigma_8, n_{\rm
  s}, h\} = \{$0.238, 0.0418, 
0.762, 0.74, 0.951, 0.73$\}$, which are consistent with the WMAP 5-year
data \citep{Komatsu2008Wmap5}.\footnote{The most notable difference is in
  $\sigma_8$, which is 1.6$\,\sigma$ lower in WMAP3 than in WMAP5.} The
primordial baryonic mass fraction of helium is assumed to be 0.248.

\subsubsection{Radiative cooling and heating}

Radiative cooling is central to simulations of the formation of
galaxies as it enables baryons to dissipate their binding energy which
allows their collapse to proceed within virialized
structures. Photo-heating by the ionizing background radiation also
plays a key role because it strongly increases pressure forces in
low-density gas, thereby smoothing out small-scale baryonic
structures.  

Previous cosmological simulations have typically included radiative
cooling assuming primordial abundances
\citep[e.g.][]{Springel2003SFH}. Some recent studies have included
metal-line cooling \citep[e.g.][]{Scannapieco2005Metals,Romeo2006ICMmetals,Oppenheimer2006Wpot,Tornatore2007ICMmetals,Choi2009Metalcooling}, 
but under the assumption of collisional ionization equilibrium and
fixed relative abundances. Photo-ionization by the UV
background radiation does not only provide a source of heat,
it also reduces the cooling rates for both primordial
and metal-enriched plasmas \citep{Efstathiou1992Cooling,Wiersma2009Cooling}.
\citet{Wiersma2009Cooling} emphasized the importance of including this
effect as well as variations in the relative abundances of the
elements.

We implemented radiative cooling and heating using the method and
tables of
\citet{Wiersma2009Cooling}\footnote{We used their equation (3) rather
  than (4) and 
  {\textsc cloudy} version 05.07 rather than 07.02.}. In brief, net
radiative cooling rates
are computed element-by-element in the presence of the cosmic
microwave background and the \citet{Haardt2001UVB} model for the UV
and X-ray background radiation
from quasars and galaxies. Hence, variations in relative abundances
and photo-ionization of heavy elements are both taken into
account. The contributions of the eleven elements 
hydrogen, helium, carbon, nitrogen, oxygen, neon, magnesium,
silicon, sulphur, calcium, and iron are interpolated as a function of
density, temperature, and redshift from tables that have been precomputed
using the publicly available photo-ionization package
{\textsc CLOUDY}, last described by \citet{Ferland1998Cloudy}, assuming the gas
to be optically thin and in (photo-)ionization equilibrium.  

The simulations model hydrogen reionization by \lq switching on\rq\, the
\citet{Haardt2001UVB} background at $z=9$. Prior to
reionization the cooling rates are computed in the presence of the
cosmic microwave background
and a photo-dissociating background which we obtain by cutting off the
$z = 9$ \citet{Haardt2001UVB} spectrum at 1~Ryd. Note that the
presence of a photo-dissociating 
background suppresses H$_2$ cooling at all redshifts. Reionization has
the effect of 
rapidly heating all of the gas to temperatures $\sim 10^4\,\K$. The
assumption that the gas is optically thin is likely to lead 
to an underestimate of the gas temperature shortly after reionization
\citep[e.g.][]{Abel1999Photoheating}. For 
the case of helium reionization we correct for this effect by 
heating the gas by a total amount
of 2~eV per atom. This extra helium reionization heating takes place
at a central redshift of 3.5, with the heating spread with a Gaussian
filter with $\sigma(z)=0.5$ in redshift. This prescription was
chosen to match observations of the temperature history of the IGM
\citep{Schaye2000IGMTemp} as shown in Fig.~1 of
\citet{Wiersma2009Chemo}. 

\subsubsection{Star formation}
\label{sec:Ref_SF}

Cosmological simulations such as ours miss both the resolution and
the physics to model 
the cold interstellar medium (ISM), let alone the formation of stars within
molecular clouds. Star formation is therefore implemented
stochastically by converting gas particles into collisionless star
particles,  which
represent simple (or single) stellar populations. We convert entire
particles, because the spawning of 
multiple star particles per gas particle affects the efficiency of
feedback from SF \citep{DallaVecchia2008Winds}. Hence, the particle
number is conserved in our simulations.
 
Gas with densities exceeding the critical density for the onset of
the thermo-gravitational instability (hydrogen number densities $n_{{\rm
    H}}=10^{-2}-10^{-1}$ cm$^{-3}$) is expected to be multiphase
and to form stars \citep{Schaye2004SF}. We therefore
impose an effective equation of state 
(EOS) with pressure $P\propto \rho^{\gamma_{{\rm eff}}}$ for
densities\footnote{Gas particles are only placed on the EOS if their
  temperature was below $10^5\,\K$ when they crossed the density
  threshold and if their density exceeds 57.7 times the cosmic
  mean. These criteria prevent SF in intracluster gas and in
  intergalactic gas at very high redshift, respectively
  \citep{Schaye2008SF}.} 
$n_{{\rm H}} >n_{{\rm H}}^*$ where $n_{{\rm H}}^*=0.1$~cm$^{-3}$,
normalised to $P/k=1.08\times 10^3\,\cm^{-3}\,\K$ at the threshold.  We use
$\gamma_{{\rm eff}}=4/3$, for which both the Jeans mass and the ratio
of the Jeans length to the SPH kernel are independent of the density,
thus preventing spurious fragmentation due to a lack of numerical
resolution \citep{Schaye2008SF,Robertson2008SFlaw}. Only gas on the
EOS is allowed to 
form stars. \citet{Schaye2008SF} demonstrated that our choice of
threshold reproduces the threshold surface density for
H$\alpha$ emission from SF that is observed in nearby galaxies.

Previous cosmological simulations used 
Schmidt-type (i.e., power-laws of the volume density) SF
laws and tuned one or more free parameters to fit the observed
Kennicutt-Schmidt SF law, which is a surface density
law. This approach is unsatisfactory as the parameters would really
need to be re-tuned if disk scale heights change as a result of
varying abundances (and hence cooling rates), SN feedback or changes
in the assumed EOS of the ISM. 

\citet{Schaye2008SF} showed that because the surface density in a
self-gravitating system is directly related to the pressure, the
Kennicutt-Schmidt SF law can be rewritten as a pressure law. This
enables one to reproduce arbitrary input Kennicutt-Schmidt laws
independently of the assumed EOS. Moreover, because the parameters are
observables, no tuning is required. Following \citet{Schaye2008SF}, we
thus compute the SFR for star-forming gas particles using
\begin{equation}
\dot{m}_\ast = m_{\rm g} A \left (1~\Msun\,\pc^{-2}\right
)^{-n} \left ({\gamma \over G} f_{\rm g} P\right )^{(n-1)/2},
\end{equation}
where $m_{\rm g}$ is the mass of the gas particle, $\gamma=5/3$ is the
ratio of specific heats (not to be confused with the effective EOS
imposed onto the ISM), $f_{\rm g}$ is the mass fraction in gas (which
we assume to be unity) and $P$ is the total pressure. The parameters
$A$ and $n$ are, respectively, the amplitude and slope of the observed 
\citet{Kennicutt1998Law} law,
$\dot{\Sigma}_\ast = A (\Sigma_{\rm g}/1~\Msun\,\pc^{-2})^n$ with
$A=1.515\times10^{-4}~\Msolyrkpcsq$ and $n=1.4$. The amplitude of this
relation has been renormalised by a factor\footnote{This 
normalization factor is calculated from the asymptotic ratio (which is
reached after only $10^8\,\yr$) of the
numbers of ionizing photons predicted from models of stellar
populations with a constant SFR \citep{Bruzual2003Synthesis}.} 1/1.65 to account for the fact
that the original analysis of \citet{Kennicutt1998Law} assumed the
\citet{Salpeter1955IMF} IMF whereas we use the
\citet{Chabrier2003IMF} IMF.   

\subsubsection{Stellar evolution and chemodynamics}
\label{sec:Ref_chemo}

Our implementation of stellar evolution and chemical enrichment is
discussed in detail in \citet{Wiersma2009Chemo}. Here, we will 
provide only a brief summary.

Each star particle represents a single stellar population that is
specified by its initial mass, age, and its chemical
composition (which it inherits from its progenitor gas particle). We
follow the timed release, by both massive
stars (Type II SNe and stellar winds) and intermediate mass
stars (Type Ia SNe and asymptotic giant branch (AGB) stars),
of all 11 elements that  
contribute significantly to the radiative cooling rates. During
each time-step, star particles distribute the mass they eject over
their neighboring gas particles\footnote{As
  discussed in \citet{Wiersma2009Chemo}, we do not change the entropy
  of the receiving particles.} using the SPH interpolation scheme. 

We assume a \citet{Chabrier2003IMF} IMF spanning the range
0.1 to 100~$\Msun$ and use the metallicity-dependent stellar lifetimes of
\citet{Portinari1998Chemo} and the complete set of nucleosynthetic yields of
\citet{Marigo2001AGByields} and \citet{Portinari1998Chemo} along with
the SN Type Ia (SNIa) yields of the W7 model of
\citet{Thielemann2003SNIayields}.  
Since SNIa are thought to result from binary evolution, a single
stellar population will
produce SNIa over an extended period. We 
implement the release of mass and energy (which we inject in thermal form)
by SNIa using empirically derived rates normalised 
to the observed cosmic SNIa rate (see Fig.~A6 of
\citealt{Wiersma2009Chemo}). For reference, our assumed 
rate implies that a fraction of about 0.025 of stars with initial mass
between 3 and 8 solar masses end their lives as Type Ia SNe. 

For the purpose of both radiative cooling and stellar evolution, we
define the abundance of a particular element as the ratio of the SPH
estimates of its mass density and the total gas mass density.
\citet{Wiersma2009Chemo} showed that the use of such `smoothed
abundances' for the cooling rates significantly increases the SFR compared to
the standard approach of using `particle metallicities', i.e., the
ratio of the elemental mass to the total mass of a particle. While
the use of smoothed abundances reduces the effects of the lack of metal
mixing inherent to SPH, it does not solve the problem.  

\subsubsection{Energy feedback from core collapse supernovae}
\label{sec:Ref_sn}

The use of an effective EOS with a pressure that exceeds that of the warm,
neutral ISM can be considered a form of weak feedback
that reflects 
the fact that energy injected by massive stars and SNe
drives small-scale turbulence. However, as we will show explicitly in
section~\ref{sec:ism}, this form of feedback does not lead to a
significant suppression of SF. Indeed, observations show
that starburst galaxies drive large-scale winds
\citep[e.g.][]{Veilleux2005windsreview} which may, over time,
eject large amounts of gas and may therefore dramatically reduce
the SFR.

As discussed in \citet{DallaVecchia2008Winds}, thermal energy from
SNe is quickly radiated away in simulations like
ours because the ratio of the heated mass to that of the star particle
is too large. In \citet{DallaVecchia2009Winds} and 
section~\ref{sec:thermal} we show that
it is possible to overcome this overcooling problem, without ad-hoc
suppression of the radiative cooling rates, by decreasing this
ratio. However, for our reference model we use the more standard
approach of injecting SN energy in kinetic form using the
prescription of \citet{DallaVecchia2008Winds}, which is a
variation of the recipe of \citet{Springel2003Multiphase}. 

After a short delay of 30~Myr, corresponding to the maximum lifetime
of stars that 
end their lives as core-collapse SNe, newly-formed star particles
inject kinetic energy into their surroundings by
kicking a fraction of their SPH neighbours in random directions. Wind
particles are not allowed to form stars for a period of $15$~Myr in
order to
avoid high velocity star particle ejection (a numerical artifact we
observed occasionally in high-resolution simulations of isolated
galaxies). These time delays are not important for the results
presented here.  

Each SPH neighbour $i$ of a newly-formed star particle $j$
has a probability of $\eta m_j/\sum_{i=1}^{N_{\rm ngb}}m_i$ of
receiving a kick with a velocity $v_{\rm w}$. Thus, if all baryonic
particles had equal mass, each newly formed star particle would kick,
on average, $\eta$ of its neighbours.
Our reference model
uses the default parameters of \citet{DallaVecchia2008Winds},
i.e.\ $\eta = 2$ and $v_{\rm w} = 600~\kms$  
Assuming that each star with
initial mass in the range $6-100~\Msun$ injects
$10^{51}\,\erg$ of kinetic energy, these parameter values imply that the
total wind energy accounts for 40 per cent of the available kinetic
energy for our IMF (if we ignore the electron capture SNe predicted by models
with convective overshoot \citep[e.g.][]{Chiosi1992Overshoot} and consider only
stars in the mass range $8-100~\Msun$, this works out to be 60 per
cent). The value $\eta=2$ was 
chosen in part because it 
roughly reproduces the peak in the cosmic SFR, as we will show later. 

Note
that contrary to the widely-used kinetic 
feedback recipe of \citet{Springel2003Multiphase}, the kinetic energy
is injected 
\emph{locally} and the wind particles are \emph{not} decoupled
hydrodynamically. As 
discussed by \citet{DallaVecchia2008Winds} and as we will show in
Section~\ref{sec:hydrodec}, these differences have important 
consequences. 

\subsubsection{Black hole growth and feedback from AGN}

The reference model does not include a prescription for the growth of
supermassive BHs and feedback from AGN. However, AGN are included in
the OWLS project as an 
optional model that is switched on for a subset of the simulations
(Sec.~\ref{sec:agn}). 

\subsection{Convergence tests}
\label{sec:convergence}

Before trying to interpret the results of numerical simulations, we
must check whether they have converged numerically. For cosmological
simulations this means checking whether the box size was sufficiently
large and whether the resolution was sufficiently high. To isolate the
effects of the size of the simulation volume and the resolution, it is
necessary to vary one while holding the other fixed. 

To test for numerical convergence we have run a suite of simulations
of the reference model. The main numerical parameters of these runs
are listed in Table~\ref{tbl:ref_sims}. The simulation names contain
strings of the form \emph{LxxxNyyy}, where \emph{xxx} is the
simulation box size in comoving $\hMpc$ and \emph{yyy} is the cube
root of the number of particles per species (dark matter or
baryonic).

For readers
not interested in the details, we first give the conclusions so that
they can skip to section~\ref{sec:variations}. Even our
25~$\hMpc$ boxes are sufficiently large to obtain a converged
prediction for the cosmic SFH. The resolution of the
\emph{L025N512} and \emph{L100N512} runs suffices for redshifts $z< 7$
and $z< 3$, respectively. While the
SFR typically increases if the resolution is improved, the situation
reverses at low redshift before convergence has been attained.

The discussion below is in parts similar to the one we presented in
\citet{Wiersma2009Chemo}, where we considered the convergence of the
cosmic metal distribution in the reference simulations. 

\subsubsection{Box size}

\begin{figure}
\resizebox{\colwidth}{!}{\includegraphics{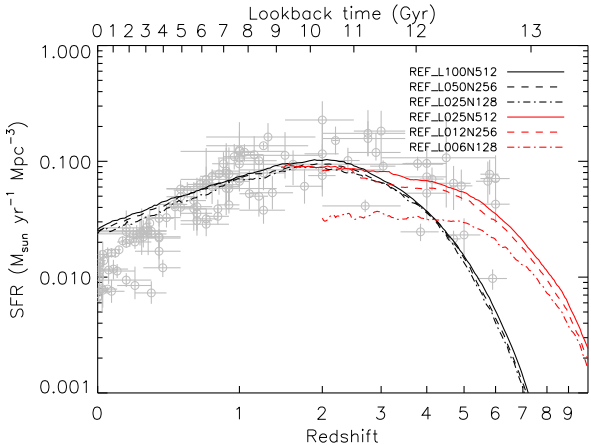}}
\caption{The effect of the box size on the cosmic SFH. The curves show
  the cosmic SFR density as a function of redshift (bottom $x$-axis)
  and lookback time (top $x$-axis) for different simulations of the
  reference model. The data points show the compilation of
  observations from \protect\citet{Hopkins2006SFH}, converted to our
  IMF and cosmology. The black curves extending down to $z=0$ are for
  simulations that use the
  same numerical resolution as model \emph{REF\_L100N512}. The red
  curves, which do not continue to $z=0$, correspond to runs with the
  resolution of \emph{REF\_L025N512}. A 25~$\hMpc$ box is sufficiently
  large to obtain a converged prediction for the SFH 
  down to $z=0$.}  
\label{fig:box}
\end{figure}

The results of cosmological simulations may depend on the size of the
simulation volume for at least two reasons. First, because the mean density in
the box is fixed by the cosmological parameters, the box size
determines what types of objects can be sampled. The distribution of
density fluctuations can only be modeled correctly on scales that are
much smaller than the box. Second, because the Fourier modes of 
the density field only evolve independently in the linear regime, the
box must be large compared to the scales on which the density contrast
is non-linear. Otherwise the missing power on scales greater than the
box size will decrease the power on the scales that are sampled by the
simulation. 

Fig.~\ref{fig:box} shows the SFR per unit comoving volume as a
function of redshift for two 
sets of simulations of the reference model. The solid curves show our
two fiducial box sizes and particle numbers: 100 comoving $\hMpc$
(\emph{REF\_L100N512}, black) and 25 comoving $\hMpc$
(\emph{REF\_L025N512}, red), both using $2\times512^3$ 
particles. The data points show the observations as compiled by
\citet{Hopkins2006SFH}. To facilitate easy comparisons, we will show
these data points and at least one of the two fiducial runs in all
subsequent figures. 

We caution the reader that the data
are subject to large systematic uncertainties due to for example the
assumed IMF and dust correction. Observe also that the scatter is
clearly too large compared with the error bars, despite 
the fact that \citet{Hopkins2006SFH} applied a uniform dust correction
and that the same IMF was assumed for all observations. Given these
uncertainties, models whose predictions are discrepant with respect to
these observations cannot automatically be ruled out. 

Focusing first on the \emph{L100} run (black, solid curve) we see a sharp
rise at high redshift, a peak at $z\approx 2$ followed by a steady
decline to $z=0$. Qualitatively this matches the data, although the
simulations appear to underestimate the SFR beyond the peak as well as
the steepness of the decline below $z=1$. Note that the height of the
peak, i.e.\ the maximum SFR, is sensitive to the fraction of the SN
energy that is used to generate galactic winds. While we fixed the
wind velocity to $600~\kms$ based on other considerations
\citep[see][]{DallaVecchia2008Winds}, the mass 
loading $\eta=2$ (which corresponds to 40 percent of the SN energy for
$v_{\rm w} =600~\kms$), was chosen partly because it gives roughly the
right maximum SFR. We will show later that the underestimate of the
SFR at $z>3$ can be attributed to a lack of numerical resolution,
while the overestimate of the SFR at $z < 0.5$ reflects the fact that
our galactic winds cannot suppress SF in massive galaxies. 

Comparing the three black curves extending to $z=0$, which correspond
to box sizes of 25, 
50, and 100~$\hMpc$, we see that even a 25~$\hMpc$ box is large enough
to obtain a converged estimate for the cosmic SFH. This is perhaps 
surprising, as there clearly exist structures with sizes that are of
the same order or greater than
this. Apparently, rare objects like clusters of galaxies do not
contribute significantly to the mean SFR. This is consistent with
\citet{Crain2009Gimic}, who used zoomed simulations to show that while
the SFR in different 25~$\hMpc$ regions varies by up to an order of magnitude,
the SFH in a region of this size whose mean density equals the cosmic
mean closely tracks the global SFH.

Comparing the three red curves that end at higher redshifts, which
correspond to box sizes of 
6.125, 12.5 and 25~$\hMpc$ and particle masses that are 64 times
smaller than those used for the black curves, we see that while a
12.5~$\hMpc$ box is nearly sufficiently large for $z>2$ (recall that
we already established that 25~$\hMpc$ is sufficiently 
large using the lower resolution simulations), 6.125~$\hMpc$ is
clearly insufficient to obtain a converged estimate of the SFH.  

Now that we have established that our fiducial box sizes are
sufficiently large, we turn our attention to the convergence with
respect to resolution.

\subsubsection{Numerical resolution}

\begin{figure*}
\resizebox{\colwidth}{!}{\includegraphics{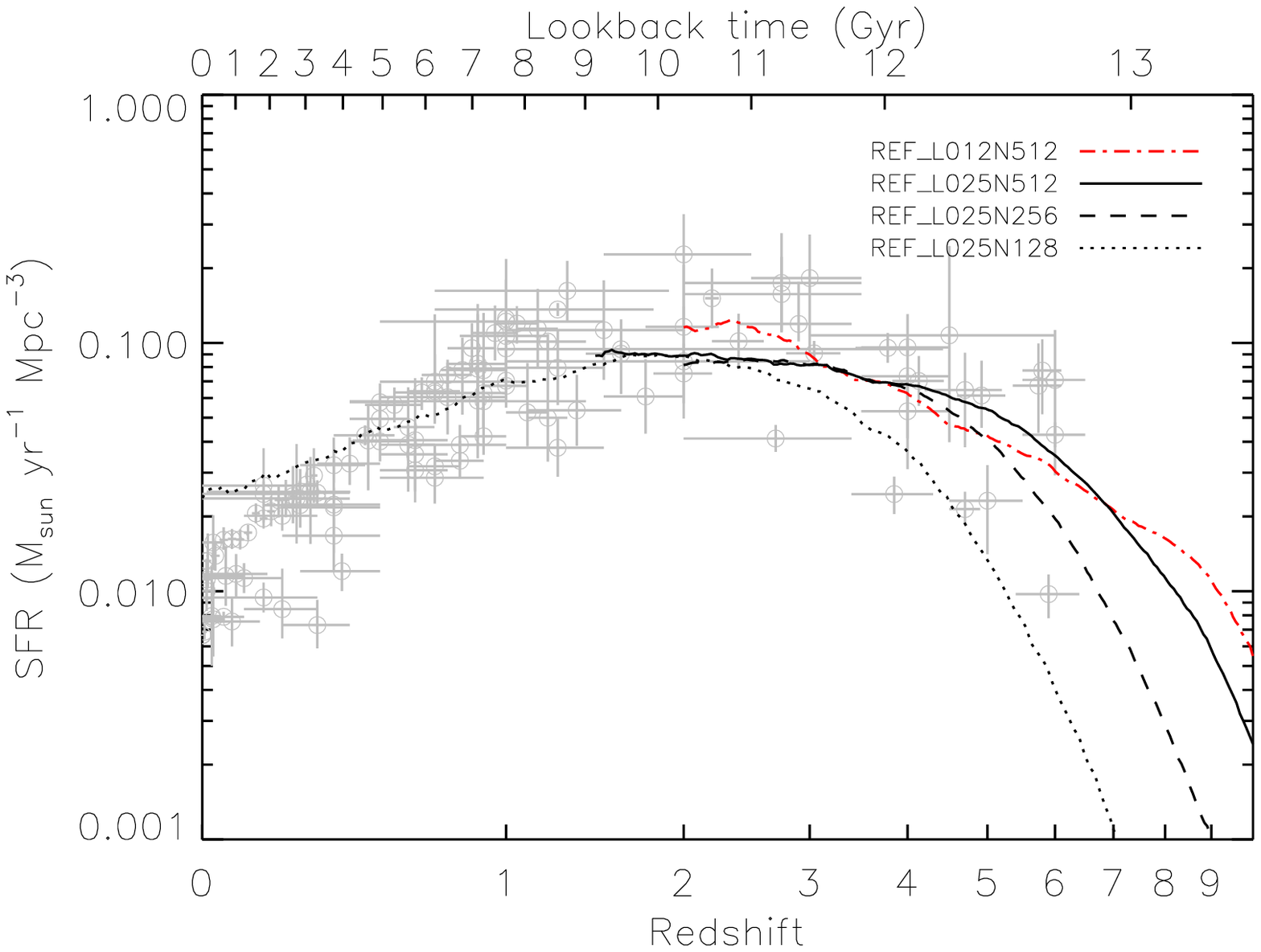}}
\resizebox{\colwidth}{!}{\includegraphics{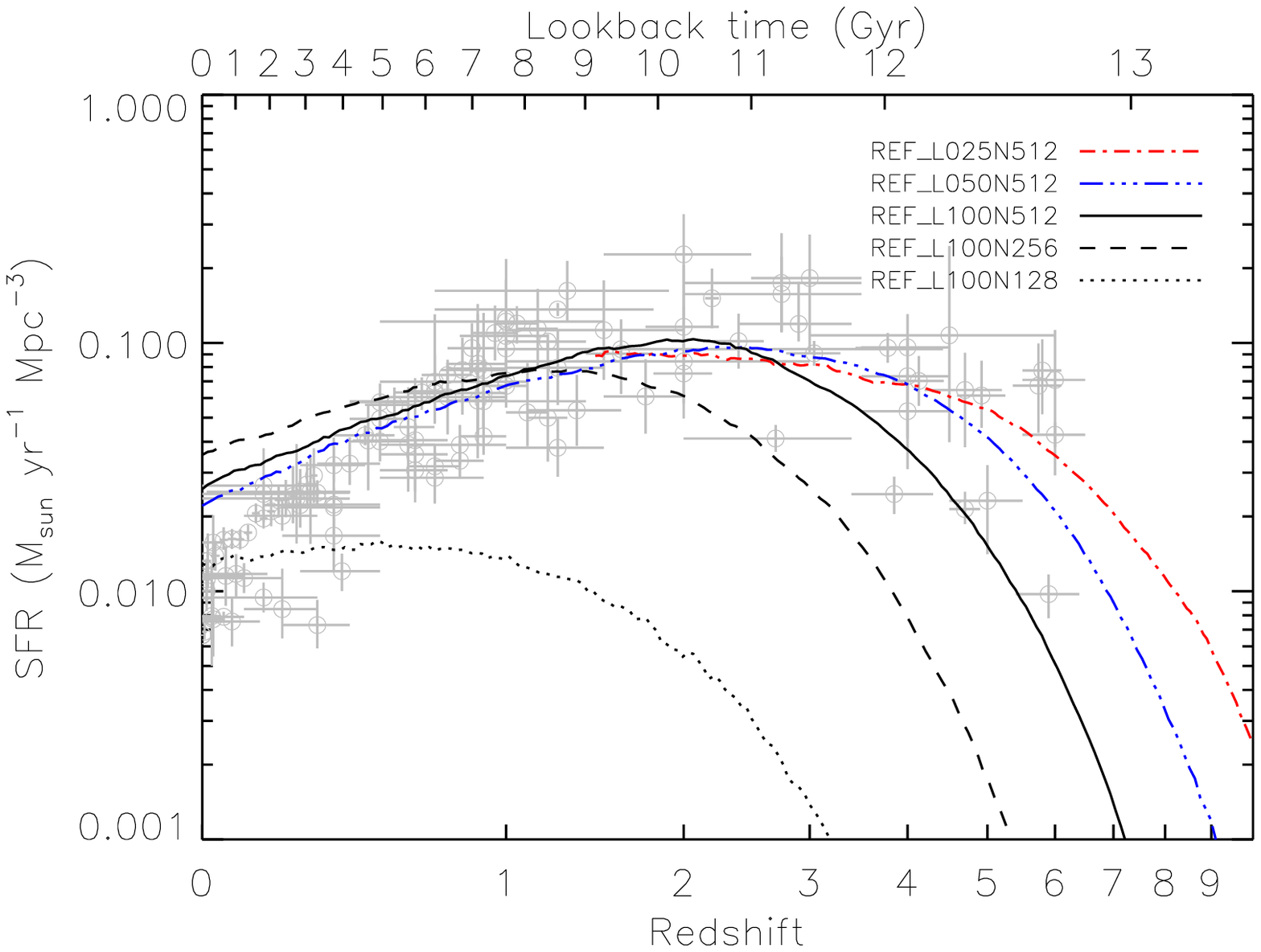}}
\caption{As Fig.~\protect\ref{fig:box}, but comparing the SFHs for
  simulations of the reference model that differ in terms of
  their numerical resolution. The left and right panels test the
  convergence of the SFH predicted by the \emph{L025N512} and
  \emph{L100N512} runs, 
  respectively. The SFH in the \emph{L025N512} run is clearly fully
  converged 
  for $z< 4$ (compare with the lower resolution \emph{L025N256}), but
  comparison with the higher resolution \emph{L012N512} (which, however, has a
  box size that is slightly too small to be converged; see
  Fig.~\protect\ref{fig:box}) suggests that it 
  is nearly converged for $z<7$. Comparing \emph{L100N512} with
  \emph{L050N512} (which uses a box size sufficiently large to provide
  a converged result; see Fig.~\protect\ref{fig:box}), we see that the
  former has nearly converged for $z<3$. Interestingly,
  increasing the resolution 
  beyond that of \emph{L100N256} \emph{decreases} the SFR at late
  times.}.    
\label{fig:resol}
\end{figure*}

While the simulation box limits the maximum sizes of the
structures that can form, numerical resolution may even affect the
properties of common objects. For example, hydrodynamical simulations
that do not resolve the Jeans scales may underestimate the fraction of
mass in collapsed structures and thus the SFR. Moreover, we cannot
expect to form dark matter haloes whose masses are comparable to or
smaller than the particle mass. 

Before showing the results of the convergence tests, it is useful to
consider what to expect. The \emph{L025N512} model has a dark matter
particle mass of $m_{\rm dm}\approx 6\times 10^6\,\hMsun$. From a
comparison of the mass functions of dark matter only simulations, we
find that $\sim 10^2$ particles are needed to robustly define a
halo. Thus, we expect the halo mass function to be converged for $M>
6\times 10^8\,\hMsun$. Comparing this to the mass corresponding to the
virial temperature below which
photo-heating is expected to suppress SF, $M \sim 0.7\times
10^8\,\hMsun \left (T_{\rm vir}/10^4\,\K\right )^{3/2} \left ((1+z)/10\right
)^{-3/2}$, we see that even after reionization we are missing haloes
expected to form stars. However, haloes of such low mass are only
expected to dominate the cosmic SFR at very high
redshift. The particle mass for \emph{L100N512} is 64 times greater
and we can thus only probe the mass function down to $M\sim 4\times
10^{10}\,\hMsun$. This will lead us to underestimate the SFR at $z\ga
3$ \citep[see e.g.\ Fig.~5 of][]{Crain2009Gimic}. 

Let us now consider the hydrodynamics. The Jeans scales depend on the
density and the temperature of the gas.
The temperature of substantially overdense\footnote{Gas with very low
overdensities can have temperatures substantially below $10^4\,\K$
due to the adiabatic Hubble expansion, but the Jeans scales corresponding to
these low densities are nevertheless large.} gas is $\ga 10^4\,\K$ in our
simulations. In 
reality, gas at interstellar densities 
($n_{\rm H}\ga 10^{-1}\,\cm^{-3}$) is sufficiently dense and
self-shielded to form a cold ($T\ll 
10^4\,\K$) gas phase, allowing it to
form stars \citep{Schaye2004SF}. However, our
simulations impose an effective EOS for gas with
densities exceeding our SF threshold of $n_{\rm H} =
10^{-1}\,\cm^{-3}$. For
our EOS ($P\propto \rho^{4/3}$) the Jeans mass is
independent of the density. Hence, if we resolve the Jeans mass
at the SF threshold, we resolve it everywhere. The Jeans
mass is given by
\begin{equation}
M_{\rm J} \approx 1 \times 10^7\,\hMsun f_{\rm g}^{3/2} \left
({n_{\rm H} \over   10^{-1}\,\cm^{-3}}\right )^{-1/2} \left ({T \over
  10^4\,\K}\right )^{3/2},
\end{equation}
where $f_{\rm g}$ is the (local) fraction of the mass in gas. Thus, we
do not expect 
convergence unless the gas particle mass $m_{\rm g} \ll
10^7\,M_{\odot}$. To achieve convergence, a simulation will, however,
also need to resolve the Jeans length $L_{\rm J}$. This implies that
the maximum, proper gravitational softening, $\epsilon_{\rm prop}$, must be
small compared with the Jeans Length 
\begin{equation}
L_{\rm J} \approx 1.5~\hkpc ~f_{\rm g}^{1/2} \left
({n_{\rm H} \over 10^{-1}\,\cm^{-3}}\right )^{-1/2} \left ({T \over
  10^4\,\K}\right )^{1/2}.
\end{equation}
Note that since $L_{\rm J}$ scales as $L_{\rm J}\propto \rho^{-1/3}$
for our EOS, $\epsilon_{\rm prop}$ will always exceed
$L_{\rm J}$ for sufficiently high densities. However, since the Jeans
mass does not decrease with density, we do not expect runaway collapse
for star-forming gas. 

Comparing the above equations with the gas particle
mass and softening scales for our fiducial simulations (see
Table~\ref{tbl:ref_sims}),  
we see that while \emph{L025N512}
marginally resolves the Jeans scales for $f_{\rm g}\approx 1$, this is not
the case for the simulations that go down to $z=0$, although \emph{L050N512}
has $m_{\rm g}\approx M_{\rm J}$ and $\epsilon_{\rm prop} \approx
L_{\rm J}$ and is therefore not far off. 
Note, however, that none of our simulations come close to resolving
the Jeans scales prior to reionization, when SF in haloes
with virial temperatures less than $10^4\,\K$ may have been important.

Fig.~\ref{fig:resol} compares the SFHs predicted by simulations
with varying resolutions. The left and right panels test the
convergence of the \emph{L025N512} and \emph{L100N512} models,
respectively. Focusing first on the solid and dashed black curves in
the left panel, we see that a particle 
mass 8 times greater than our fiducial value (and a softening twice
our fiducial value) is sufficient for $z< 4$. For $z<2$ even a
particle mass that is a factor 64 smaller appears to be
sufficient. Comparison of our fiducial run with the higher resolution
run \emph{L012N512} (red, dot-dashed), indicates that the former is
likely nearly 
converged for $z < 7$. Note that we do not
expect perfect agreement when comparing \emph{L025} and
\emph{L012} runs even if they have converged in terms of resolution
since the two runs necessarily have different initial conditions and a
12.5~$\hMpc$ box is not fully converged in terms of box size (see
Fig.~\ref{fig:box}).  

The kink at $z=9$ in the SFH of \emph{L012N512} is due to the negative feedback
associated with reheating during reionization. While this effect is
easily visible for \emph{L012N512}, it is only just detectable in
\emph{L025N512}. This is expected. Reionization will suppress
SF in haloes with virial temperatures $\la 10^4\,\K$, which corresponds
to halo masses of $\sim 10^8\,\Msun$ at this redshift. Such haloes are
resolved with $\sim 10^2$ particles in \emph{L012N512}, but contain
only $\sim 10$ particles in \emph{L025N512}.

Comparing the three black curves in the right panel, which show the
SFH for runs \emph{L100N128}, \emph{L100N256}, and \emph{L100N512}, we see
no evidence for full convergence, although the difference between the two
highest resolution runs is small for $z<1.5$. Indeed, comparison with
\emph{L050N512} and \emph{L025N512}, whose box sizes we have already
shown to be 
sufficiently large, reveals that the fiducial run \emph{L100N512} has
nearly converged for $z< 3$ (and \emph{L050N512} for $z\la 4$).  

An increase in the numerical resolution typically increases the SFR,
particularly at high redshift, when it is dominated by haloes near the
resolution limit. Observe, however, that the opposite happens at low
redshift once the resolution is increased beyond that of
\emph{L100N256}. This reduction reflects
the fact that, in simulations with higher resolution, the gas that
would otherwise be available or SF has already been used up or
ejected by lower mass progenitors at higher redshifts. 

Single simulations currently lack the dynamic range to
obtain a converged result for the SFH over a wide range of
redshifts. One strategy that has been used to overcome this limitation
\citep[e.g.][]{Springel2003SFH} is to combine a suite of simulations
with different box sizes and to use, for each redshift, the one that
yields the highest SFR. This procedure comes down to plotting all SFH
curves and using the envelope that encompasses all as the best
estimate for the SFH. However, our finding that, at low redshift, the
predicted SFR decreases as it approaches convergence, indicates that
this procedure may overestimate the SFR at late times. Because the
low-redshift SFR depends on the amount of gas that was consumed or
ejected at earlier times, one really does need a large dynamic range
to model the SFH down to $z=0$.

Summarizing, the convergence tests are consistent with our
expectations based on our estimates of the minimum resolved dark halo
mass and a comparison of the mass and length resolutions with the Jeans
scales. For our purposes, the resolution of the \emph{L100N512} run
suffices for $z< 3$ and that of \emph{L025N512} for
$z\la 7$. 

Before investigating the physics driving the predicted SFH, we
caution the reader that convergence of our reference model does not
automatically imply that other physics variations are also converged 
at the same resolution. On the other hand, it would be surprising if resolution
effects would change qualitative conclusions drawn from comparisons in
the regime for which \emph{REF} is converged. This situation would
change, however, if we did not impose an effective equation of state
onto the ISM because in that case the Jeans scales could become much
smaller than in the models considered here.

\section{Variations on the reference model}
\label{sec:variations}
\begin{table*} 
\begin{center}
\caption{List of main physics variations employed in the OWLS project.
  From left to right the columns show the simulation name, whether or
  not the simulation was run in the 25~Mpc$/h$ and 100~Mpc$/h$ boxes
  respectively, the section number containing the description of
  the model, and a very brief description of the changes in the model
  relative to the \emph{REF} simulation. Except for the \emph{MILL}
  runs, all simulations that were run using the same box size used
  identical initial conditions.}  
\label{tbl:sims}
\begin{tabular}{lccll}
\hline
Simulation & L025 & L100 & Section & Description\\ 
\hline 
\emph{AGN} & $\surd$ & $\surd$&\ref{sec:agn} & Includes AGN\\
\emph{DBLIMFCONTSFV1618} & $\surd$ & $\surd$ & \ref{sec:dblimf} & Top-heavy IMF at high pressure, cont.\ SF law, extra SN energy in wind velocity\\ 
\emph{DBLIMFV1618} & $\surd$ & $\surd$ & \ref{sec:dblimf} & Top-heavy IMF at high pressure, extra SN energy in wind velocity\\
\emph{DBLIMFCONTSFML14} & $\surd$ & $\surd$ & \ref{sec:dblimf} & Top-heavy IMF at high pressure, cont.\ SF law, extra SN energy in mass loading\\
\emph{DBLIMFML14} & $\surd$ & $\surd$ & \ref{sec:dblimf} & Top-heavy IMF at high pressure, extra SN energy in mass loading\\
\emph{EOS1p0} & $\surd$ & $\surd$ & \ref{sec:ism} & Slope of the effective EOS changed to $\gamma_{\rm eff}=1$\\
\emph{EOS1p67} & $\surd$ & - & \ref{sec:ism} & Slope of the effective EOS changed to $\gamma_{\rm eff}=5/3$\\
\emph{IMFSALP} & $\surd$ & $\surd$ & \ref{sec:imfsalp} & Salpeter (1955) IMF\\
\emph{IMFSALPML1} & $\surd$ & -  & \ref{sec:imfsalp} & Salpeter (1955) IMF; wind mass loading $\eta=2/1.65$\\
\emph{MILL} & $\surd$ & $\surd$ & \ref{sec:cosmology} & Millennium
simulation cosmology, 
$\eta=4$ (twice the SN energy of \emph{REF})\\
\emph{NOAGB\_NOSNIa} & - & $\surd$ & \ref{sec:chemo} & No mass loss
from AGB stars and SNIa\\
\emph{NOHeHEAT} & $\surd$ & - & \ref{sec:reionization} & No extra heat input around helium reionization\\
\emph{NOREION} & $\surd$ & - & \ref{sec:reionization} & No hydrogen reionization\\
\emph{NOSN} & $\surd$ & $\surd$ & \ref{sec:sn} & No SN energy feedback from SNe\\
\emph{NOSN\_NOZCOOL} & $\surd$ & $\surd$ & \ref{sec:zcool0} & No SN energy feedback from SNe and cooling assumes primordial abundances\\
\emph{NOZCOOL} & $\surd$ & $\surd$ & \ref{sec:zcool0} & Cooling assumes primordial abundances\\
\emph{REF} & $\surd$ & $\surd$ & \ref{sec:reference} & Reference model\\
\emph{REIONZ06} & $\surd$ & - & \ref{sec:reionization} & Hydrogen reionization occurs at $z=6$\\
\emph{REIONZ12} & $\surd$ & - & \ref{sec:reionization} & Hydrogen reionization occurs at $z=12$\\
\emph{SFAMPLx3} & $\surd$ & - & \ref{sec:sflaw} & Normalization of
Kennicutt-Schmidt SF law increased by a factor of 3\\
\emph{SFAMPLx6} & $\surd$ & - & \ref{sec:sflaw} & Normalization of
Kennicutt-Schmidt SF law increased by a factor of 6\\
\emph{SFSLOPE1p75} & $\surd$ & - & \ref{sec:sflaw} & Slope of
Kennicutt-Schmidt SF law increased to 1.75\\
\emph{SFTHRESZ} & $\surd$ & - & \ref{sec:thresh} & Critical density for onset of SF is a function of metallicity (Eq.~\ref{eq:sfthresz})\\
\emph{SNIaGAUSS} & - & $\surd$ & \ref{sec:chemo} & Gaussian SNIa delay function\\
\emph{WDENS} & $\surd$ & $\surd$ & \ref{sec:winds_ce} & Wind mass
loading and velocity depend on gas density (SN energy as \emph{REF})\\
\emph{WHYDRODEC} & $\surd$ & - & \ref{sec:hydrodec} & Wind particles
are temporarily hydrodynamically decoupled\\ 
\emph{WML1V848} & $\surd$ & $\surd$ & \ref{sec:winds_ce} & Wind mass
loading $\eta=1$, velocity $v_{\rm w}=848\,{\rm km/s}$ (SN energy as \emph{REF})\\
\emph{WML4} & $\surd$ & $\surd$ & \ref{sec:sn} & Wind mass loading
$\eta=4$ (twice the SN energy of \emph{REF})\\
\emph{WML4V424} & $\surd$ & - & \ref{sec:winds_ce} & Wind mass loading
$\eta=4$; wind velocity $v_{\rm w}=424\,{\rm km/s}$ (SN energy as \emph{REF})\\
\emph{WML8V300} & $\surd$ & - & \ref{sec:winds_ce} & Wind mass loading
$\eta=8$; wind velocity $v_{\rm w}=300\,{\rm km/s}$ (SN energy as \emph{REF})\\
\emph{WPOT} & $\surd$ & $\surd$ & \ref{sec:wmom} & Wind mass loading
and vel.\ vary with grav.\ potential (``Momentum-driven'')\\
\emph{WPOTNOKICK} & $\surd$ & $\surd$ & \ref{sec:wmom} & Same as \emph{WPOT} except that no extra velocity kick is given to winds \\
\emph{WTHERMAL} & $\surd$ & - & \ref{sec:thermal} & SN energy injected thermally\\
\emph{WVCIRC} & $\surd$ & $\surd$ & \ref{sec:wmom} & Wind mass loading and vel.\ vary with halo circ.\ vel.\ (``Momentum-driven'')\\
\hline
\end{tabular}
\end{center}
\end{table*}

In this section we describe the full set of OWLS runs. Most
simulations differ from the reference model in only the choice of a
single parameter or the presence of a certain aspect of the subgrid
physics.  In this way, cross-comparison of different simulations
allows us to isolate the effects of different physical processes and
the importance of different numerical parameters.  The full list of
simulations is shown in Table~\ref{tbl:sims}, which lists the
simulation identifier, indicates whether or not a given simulation was run
in the 25 and 100~$\hMpc$ boxes and gives a reference to the
section that discusses the simulation. Except for the \emph{MILL}
runs, all simulations that were run in the same box size used
identical initial conditions.  

The order in which we present the different variations roughly
parallels the order in which the subgrid models were discussed in
section~\ref{sec:ref_descr}. The different subsections can be read
independently of each other. We begin by comparing our WMAP-3
cosmology to that of the WMAP-1 cosmology, which was for example
assumed in the widely used ``Millennium simulation''
\citep{Springel2005Mill}. In sections~\ref{sec:zcool0} and
\ref{sec:reionization} we investigate aspects of the radiative cooling
and heating by turning off metal-line cooling and varying the redshift
of reionization, respectively. We study the importance of our
treatment of the unresolved ISM in section~\ref{sec:ism} by varying
the EOS. We vary the subgrid model for SF in section~\ref{sec:sf},
where we try a metallicity-dependent SF threshold as well as a range
of Kennicutt-Schmidt SF laws. In section~\ref{sec:chemo} we
investigate the effect of intermediate mass stars by turning off mass
loss from AGB stars and by varying the time delay function for Type
Ia SNe. Section~\ref{sec:imf} investigates the effect of using a
Salpeter IMF and an IMF that becomes top-heavy at high
pressures. Many aspects of our prescription for kinetic SN feedback are
varied in section~\ref{sec:sn}. We not only try a range of parameter
values, but also check the effect of temporarily
decoupling the hydrodynamical forces on wind particles. This section
also investigates a promising way of injecting SN feedback in
thermal form. Another form of feedback from young stars is studied in
section~\ref{sec:wmom}, where we discuss the results of 
various simplified implementations of radiatively driven
winds. Finally, we study the effect of AGN feedback in
section~\ref{sec:agn}. 

\subsection{Cosmology}
\label{sec:cosmology}

\begin{figure}
\resizebox{\colwidth}{!}{\includegraphics{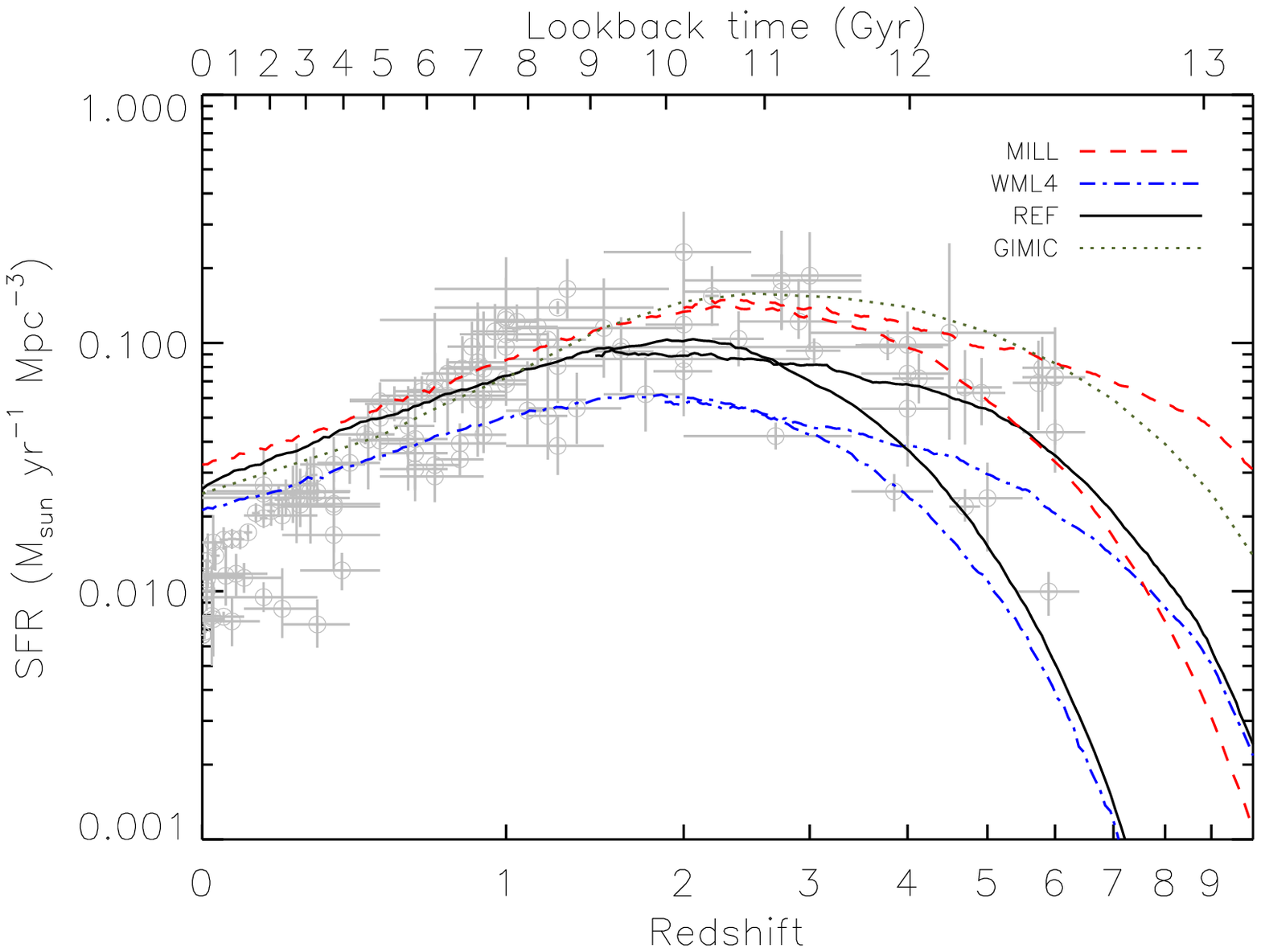}}
\caption{The effect of cosmology on the cosmic SFH. The curves show
  the cosmic SFR density as a function of redshift (bottom $x$-axis)
  and lookback time (top $x$-axis) for models \emph{MILL} (red, dashed) and
\emph{WML4} (blue, dot-dashed). For comparison, the results for the
reference model are also shown (black, solid). 
Results are shown for both the 25 and
the 100~$\hMpc$ boxes, with the smaller box predicting higher SFRs at
high redshift. All simulations used $2\times 512^3$ particles.
The data points show the compilation of
observations from \protect\citet{Hopkins2006SFH}, converted to our IMF
and our fiducial cosmology. For comparison, we also show the SFH
predicted by GIMIC \citep[][olive, dotted]{Crain2009Gimic}. 
Models \emph{MILL} and \emph{WML4} differ only in terms of
the assumed cosmology. In particular, model \emph{MILL} assumes values
for $\Omega_{\rm b}$ and $\sigma_8$ that are, respectively, 8 and 22
per cent higher than our fiducial values. Both \emph{MILL} and
\emph{WML4} assume that the wind mass loading, and hence the fraction
of the SN energy that is injected in the form of winds, is twice as
high as for model \emph{REF}.  
Note that the lookback time axis (top $x$-axis) applies only to the
\emph{REF} cosmology. Although the data points assume the \emph{REF}
cosmology, they would be very similar for the \emph{MILL} cosmology.
Compared to our fiducial WMAP-3 cosmology, the
\emph{MILL} cosmology predicts higher SFRs, particularly at high
redshift.}    
\label{fig:cosmology}
\end{figure}

To investigate the dependence on cosmology and 
in order to facilitate comparisons to earlier work, we change the
cosmological parameters from our fiducial WMAP 3-year values
\citep{Spergel2007Wmap3} to the cosmology used in
many studies including the Millennium Simulation
\citep{Springel2005Mill}. We refer to this latter set of cosmological
parameters, which were chosen to be consistent with a combined
analysis of the 2-degree field galaxy redshift survey and the
first-year WMAP data, as the `Millennium cosmology' and denote the
models \emph{MILL}. The Millennium cosmology uses the cosmological
parameter values 
$\{\Omega_{\rm m},\Omega_{\rm b},\Omega_\Lambda,\sigma_8,n_{\rm
  s},h\}=\{0.25, 0.045, 
0.75, 0.9, 1.0, 0.73\}$. Note that because of the change in the values
of $\Omega_{\rm m}$ and $\Omega_{\rm b}$, the dark matter and the
(initial) baryonic particle masses are, respectively,
4.5 and 7.7 percent higher for the \emph{MILL} run than for the
\emph{REF} model.

The main differences with respect to the reference model
are the values of $\Omega_{\rm b}$ and $\sigma_8$ which are,
respectively, 8 and 22 percent higher for \emph{MILL} than for
\emph{REF}. Both changes 
are expected to increase the SFR. The higher value of $\sigma_8$
has a particularly large effect at high redshift, because structure
formation proceeds faster in the \emph{MILL} cosmology. In order
to roughly match the peak in the observed SFH, we doubled the mass loading
factor to $\eta = 4$ for the SN driven winds. Hence, the winds account
for 80 percent of the available energy from SNe. To isolate the effect
of cosmology, we therefore compare the 
\emph{MILL} simulation to model \emph{WML4} which employs the same
wind parameters, 
but is otherwise identical to the reference model. 
   
Fig.~\ref{fig:cosmology} compares the SFHs in the \emph{MILL} (dashed,
red) and \emph{WML4} (dot-dashed, blue) runs. The change from the
WMAP-3 to the \emph{MILL}
cosmology strongly boosts the SFR. The difference increases with
redshift from about
0.2~dex at $z=0$ to 0.34~dex at $z=2$ (for both box sizes). By $z=9$
the difference has increased to 1.0~dex. Clearly, for a quantitative
comparison with observations, it is important to use the correct
cosmology. At high redshift, when the haloes that dominate the SF in
the simulation correspond to rare fluctuations, the predicted cosmic SFR
becomes extremely sensitive to the value of $\sigma_8$. We will
show in Haas et al.\ (in preparation) that the differences are
much smaller for haloes of a fixed mass, which implies that the change
in the halo mass function accounts for most of the differences in the
SFHs predicted for the two cosmologies. 

The olive, dotted curve in Fig.~\ref{fig:cosmology} shows the SFH
predicted by the Galaxies-Intergalactic Medium Interaction Calculation
(GIMIC, \citealt{Crain2009Gimic}). GIMIC consists of a series of
hydrodynamical simulations that zoom in on 25~Mpc subvolumes of the
$500~\hMpc$ dark matter only Millennium simulation and was run using
the same code and parameter values as
\emph{MILL}. Fig.~\ref{fig:cosmology} shows the SFH computed from the
weighted average of the five GIMIC subvolumes. The particle mass
(gravitational softening)
used for the GIMIC runs\footnote{These are the numbers for the
  intermediate resolution GIMIC runs. The high-resolution runs use the
  same particle masses and force resolution as our $25~\hMpc$ box, but
  they end at $z=2$ and do not include the highest density subvolume.} 
is 8 (2) times larger
than that of our $25~\hMpc$ box and thus 8 (2) times smaller than for
our $100~\hMpc$ box. At very high redshift the SFR in the GIMIC run is
intermediate between that of our two \emph{MILL} box sizes. For $z<7$
it is very close to that of the $25~\hMpc$ box and at $z<3$ it agrees
with the $100~\hMpc$ run, although the GIMIC SFR falls of somewhat more steeply
below $z=2$. These differences are exactly what is expected
from resolution effects as can be seen by comparing the SFHs for the
25, 50 and $100~\hMpc$ boxes shown in the right panel of
Fig.~\ref{fig:resol}. The excellent agreement confirms that our box
sizes are sufficiently large to obtain a converged prediction for the
cosmic SFH.

\subsection{Metal-line cooling}
\label{sec:zcool0}

\begin{figure}
\resizebox{\colwidth}{!}{\includegraphics{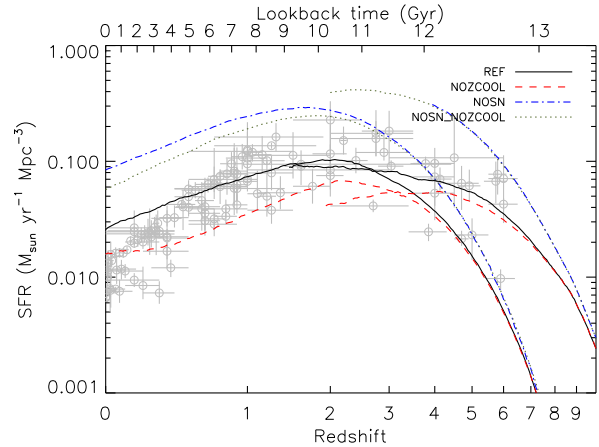}}
\caption{As Fig.~\protect\ref{fig:cosmology}, but comparing the SFHs for
  models with and without metal-line cooling, both in the presence and
absence of SN-driven winds. Except at very high redshift, metal-line
cooling strongly increases the SFR. The boost due to metal-line
cooling is greater when SN feedback is included, which implies that
metals radiate away a significant fraction of the energy injected by
SNe.}   
\label{fig:fbenergy_zcool0}
\end{figure}

Simulations without any radiative cooling are of interest for the
study of hot gas in groups of clusters of galaxies (we have
run such a simulation for this purpose in the 100~$\hMpc$ box), but in
order to form stars, the gas must be able to radiate away its binding
energy. Despite the importance of cooling, most cosmological studies
still use 
highly simplified prescriptions, ignoring metal-line cooling or
including it under the assumption of collisional ionization
equilibrium and fixed relative abundances. Our simulations are the
first to compute the cooling rates element-by-element and the first
high-resolution simulations of 
cosmological volumes that include the effect of photo-ionisation on
the heavy elements.

Fig.~\ref{fig:fbenergy_zcool0} compares the reference simulations
with runs that ignored metal-line cooling (\emph{NOZCOOL}; dashed,
red). As expected, 
the two agree at very high redshift where there has not been enough
time to enrich the gas significantly and where much of the gas falls
in cold \citep[e.g.][]{White1991GF,Birnboim2003Coldaccr,Keres2005Gasaccr}. At late times, the runs without
metal-line cooling consistently predict lower SFRs. For our
high-resolution \emph{L025N512} runs the difference increases with
cosmic time to 0.3~dex at 
$z=2$. While the SFR increases to $z=2$ for \emph{REF}, it peaks at
$z=4$ when metal-line cooling is ignored. Interestingly, for the
\emph{L100} runs 
the difference decreases after peaking at about 0.4~dex
around $z=0.4$.  

Also shown in Fig.~\ref{fig:fbenergy_zcool0} are four\footnote{Note that
\emph{NOSN\_NOZCOOL\_L025N512} was stopped earlier.} runs without
SN-driven winds (but still including metal production and mass
loss from SNe), both with and without metal-line cooling. Clearly, SN
feedback strongly suppresses the SF, a point that we will come back
to in section~\ref{sec:sn}. 

Interestingly, while metal-line
cooling also enhances the SFR in the absence of SN feedback, its effect
is smaller than when SN feedback is included. Put another way, the
factor by which SN feedback reduces the SFR is smaller when metal-line
cooling is included. There are two possible explanations for this
effect, which may both be right. First, metal-line cooling
may reduce the efficiency of SN feedback, probably because it increases
radiative losses in gas that has been shock-heated by the wind. Second, SN
feedback may increase the effect of metal-line cooling, probably because
it increases the fraction of the gas that is enriched. The former
explanation is likely to be most relevant for galaxy groups, as we
will show elsewhere that galactic winds do not dominate the enrichment
of the intragroup medium.

Recently, \citet{Choi2009Metalcooling} have also investigated the effect of
metal-line cooling on the SFH. While they also used \textsc{gadget},
their simulations used about an order of magnitude fewer particles and
were stopped at higher redshifts. They
did not include stellar evolution, they assumed a different cosmology
and used different subgrid prescriptions for SF and SN feedback. Their
cooling rates were 
taken from \citet{Sutherland1993Cooling} and assume fixed
relative abundances and collisional ionization equilibrium. Hence,
there are many differences compared to our implementation of cooling,
as we do include stellar evolution and compute the cooling rates
element-by-element, including the effects of photo-ionisation by the
meta-galactic UV/X-ray background. For their simulations with resolutions
similar to our \emph{L025N512} runs (but a much smaller box), they found that
metal-line cooling increases the SFR by less than 0.1~dex at $z=3$,
whereas we find 0.16~dex. In
simulations with resolution comparable to 
our \emph{L100N512}, they found an increase of about 0.25~dex by $z=1$
whereas we find 0.3~dex. Thus, the results are broadly consistent
although we find a somewhat larger boost due to metal cooling.

Clearly, except at very high redshift, metal-line cooling is very
important. It boosts the SFRs by allowing more of the gas that
accretes onto haloes to cool and by
reducing the efficiency of SN feedback. Without metal cooling,
predictions for the total amount of 
stars formed could easily be low by a factor of two. Note, however,
that we may have overestimated the effect of metal-line cooling for massive
galaxies, because we have not included any feedback processes capable
of stopping cooling flows in such systems. This results in an
overestimate of the SFRs and the metallicities in the central regions
of groups and clusters of galaxies. On the other hand, as discussed in
\citet{Wiersma2009Chemo}, the fact 
that SPH underestimates small-scale metal-mixing (because metals are
stuck on particles) causes us to underestimate the total mass that
has been enriched, while overestimating the metallicity of the
particles that have received metals. \cite{Wiersma2009Chemo} found
that the net effect of increased metal mixing is to boost the SFR.

\subsection{Reionization}
\label{sec:reionization}

\begin{figure}
\resizebox{\colwidth}{!}{\includegraphics{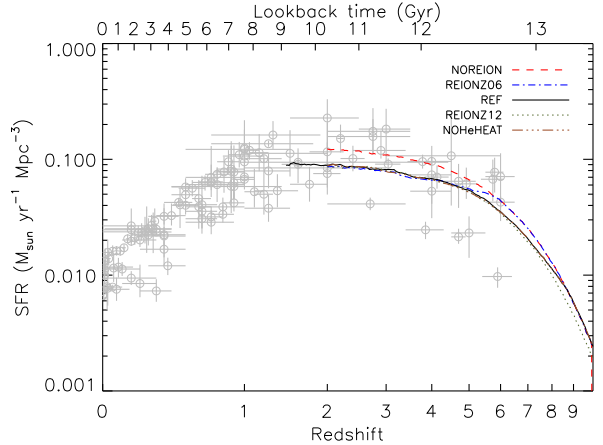}}
\caption{As Fig.~\protect\ref{fig:cosmology},
but comparing the SFHs for models in which hydrogen was reionized --- by
switching on a uniform ionizing background --- at redshift
$z_{\rm r} = 12$ (\emph{REIONZ12}), 9 (\emph{REF}), 6
(\emph{REIONZ06}), or not at all (\emph{NOREION}). All simulations use
a 25~$\hMpc$ box and $2\times 512^3$ particles. After the ionizing
background is switched on, the SFR quickly changes to the rate
predicted by the model with the highest redshift of
reionization. Note that the factor by which reionization suppresses
the SFR is limited by the resolution of the simulations and will be more
severely underestimated at higher redshifts. Turning off the extra heat
input of 2~eV per atom around helium reionization ($z\approx 3.5$,
model \emph{NOHeHEAT}) has no discernible effect on the SFH.} 
\label{fig:reion}
\end{figure}

As our simulations do not include radiative
transfer, we need to assume the background
radiation is uniform and that the gas is optically thin.
Hydrogen reionization is thus implemented in our simulations by switching
on the \citet{Haardt2001UVB} model for the ionizing background
radiation at some redshift $z_{\rm r}$, corresponding to
the epoch of reionization. Note, however, that we assume that a
photo-dissociating background is already present at $z>z_{\rm r}$,
which effectively suppresses molecular cooling at all redshifts. As described in
\citet{Wiersma2009Cooling}, switching on the ionizing radiation
results in a sudden increase in the 
radiative heating rate and a sudden decrease in the radiative cooling
rate above $10^4\,\K$. As a result, cold gas is quickly heated to $T
\sim 10^4\,\K$, removing gas from haloes with virial temperatures
$<10^4\,\K$ \citep[e.g.][]{Couchman1986Firstobjects,Okamoto2008Photoheating}.

Fig.~\ref{fig:reion} compares \emph{L025N512} runs with $z_{\rm r} = 12$, 9
(i.e.\ the reference model), 6 and 0 (i.e.\ no
reionization)\footnote{The run with $z_{\rm r} 
  = 12$ uses the $z=9$ \citet{Haardt2001UVB} model for
  $z=9-12$.}. After reionization, the
SFR deviates from the curve corresponding to the simulation
without an ionizing background (\emph{NOREION}) and quickly asymptotes
to the model with the 
highest redshift of reionization (\emph{REIONZ12}). Apparently, the gas in
the simulation rapidly loses memory of the time of reheating, as was
also found by \citet{Pawlik2009Clumping}. This is expected, as the
sound-crossing time scale is only $10^8\,\yr\left (l/1~{\rm kpc}\right
)$ for $10^4\,\K$ gas. 

Naively, one would have expected the suppression of the SFR due to
photo-ionization to decrease at late times, as haloes with $T_{\rm
  vir} \gg 10^4\,\K$ start to dominate the cosmic SFR. Interestingly,
we do not find this. If anything, the suppression keeps
increasing with time, reaching 0.15 dex (a thirty percent reduction)
by $z=2$. While this is probably mostly a resolution effect, it does
indicate that photo-ionization also reduces the SFR in
haloes with higher virial temperatures, either because of the
reduction of the cooling rates \citep{Wiersma2009Cooling} or because it
makes the cold gas more susceptible to galactic winds
\citep{Pawlik2009Photowinds}. 

We emphasize that because of our
limited resolution (Haas et al., in preparation, show that we  
underestimate the SFR in haloes with masses less than
$10^{10}\,\Msun$) and because we assume the presence of a
photo-dissociating background at all redshifts, it is likely  
that we have strongly underestimated the reduction of the SFR due to
photo-heating and that this underestimate becomes more severe at
higher redshifts. Moreover, our assumption that the gas is optically thin
results in an underestimate of the heating rates during reionization.

Helium is thought to have been reionized around $z\approx 3.5$ and the
increase in the photo-heating rates associated with this event can
explain the relatively high temperature of the IGM inferred from
observations of quasar absorption spectra
\citep[e.g.][]{Schaye2000IGMTemp}. As described in
\citet{Wiersma2009Chemo}, by injecting 2~eV per atom at $z\approx
3.5$, we are able to match the observationally inferred
temperatures. Fig.~\ref{fig:reion} shows that omitting this extra
heat (\emph{NOHeHEAT}) does not yield any noticeable changes in the
SFH. This is not surprising, as the temperature increase is confined
to low-density gas, far away from galaxies, for which adiabatic
cooling dominates over radiative cooling.

\subsection{The equation of state of the ISM}
\label{sec:ism}
  
\begin{figure}
\resizebox{\colwidth}{!}{\includegraphics{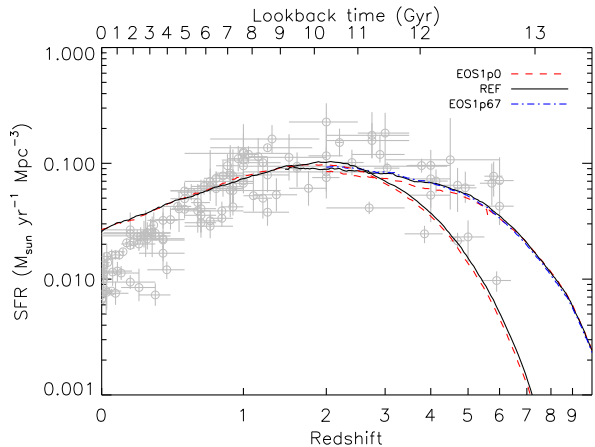}}
\caption{As Fig.~\protect\ref{fig:cosmology},
but comparing the SFHs for models assuming different equations of
state for the unresolved ISM. The slope of the polytropic EOS imposed
on the ISM is 1.0
(i.e.\ isothermal) for model \emph{EOS1p0} (red, dashed), $4/3$ for
\emph{REF} (solid, black), and 
$5/3$ (i.e.\ adiabatic) for \emph{EOS1p67} (blue, dot-dashed). The SFH
is insensitive to the assumed slope of the polytropic EOS that is
imposed onto the ISM.}  
\label{fig:eos}
\end{figure}

Our simulations have neither the resolution nor the physics to model
the multiphase ISM. As discussed in section~\ref{sec:Ref_SF}, we
therefore impose a polytropic EOS with slope $\gamma_{\rm eff} = 4/3$ for
gas with densities that exceed our SF threshold of $n_{\rm
  H}=0.1~\cm^{-3}$. This slope was 
chosen because it results in a constant Jeans mass and thus suppresses
artificial fragmentation. In this section we will check the effect of
varying the slope of the EOS. \cite{Schaye2008SF} showed that changes
in the EOS can significantly alter the morphology of galaxies. A
softer EOS results in tighter spiral arms, thinner disks, and
increased fragmentation.  

Different groups use different prescriptions for the ISM. For example,
\cite{Springel2003Multiphase} use a complicated function that results
from a semi-analytic model of the multiphase ISM. They interpret the
pressure implied by their EOS, which is steeper than 4/3 at densities
similar to our SF threshold, as a form of SN feedback.  In the
past, many cosmological simulations have 
been run that do not impose an EOS, but which also do
not include the physics necessary to model the cold interstellar gas
phase (e.g.\ radiative transfer and molecule formation). Such simulations
effectively use an isothermal EOS. 

Fig.~\ref{fig:eos} compares the SFHs of runs with $\gamma_{\rm
  eff} = 1$ (i.e.\ isothermal), 4/3 (\emph{REF}), and 5/3
(i.e.\ adiabatic). Clearly, changes in the EOS do not have a
significant effect on the predicted SFH. This may be surprising, given
that the EOS can strongly affect the structure of galaxies. 

One
reason why the results are insensitive to the EOS is that we use the
prescription for SF of \citet{Schaye2008SF}. As discussed by 
these authors, for self-gravitating systems such as galaxies, the
observed Kennicutt-Schmidt surface density law is in effect a pressure
law. 
By implementing it as a pressure law, we can thus reproduce the
observed SF law independently of the 
assumed EOS of the star-forming gas. 
Previous cosmological
simulations have, however, used volume density laws, in which case
the predicted Kennicutt-Schmidt law must depend on the assumed
EOS because the latter sets the scale height of the disk. If the EOS
is changed, then the same surface density corresponds to a different
volume density, but the relation between surface density and pressure
will remain unchanged. It is therefore not clear whether
the results of previous
simulations are as insensitive to the imposed EOS as we find
here. However, we will show in the next section that the results are
in fact insensitive to the gas consumption time scale, because
SN feedback enables galaxies to regulate their SFRs.

\subsection{Star formation}
\label{sec:sf}

\subsubsection{The star formation threshold}
\label{sec:thresh}

\begin{figure}
\resizebox{\colwidth}{!}{\includegraphics{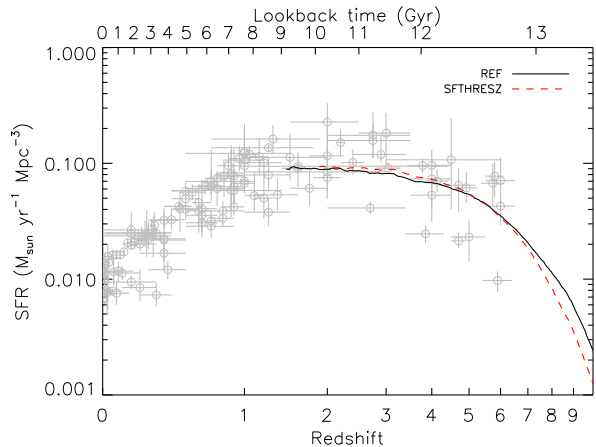}}
\caption{As Fig.~\protect\ref{fig:cosmology},
but comparing the SFHs for the reference model (black, solid), which
uses a fixed threshold density for SF, to that of model
\emph{SFTHRESZ} (red, dashed), for which the
SF threshold decreases with metallicity as predicted by
\protect\cite{Schaye2004SF}. The threshold densities in the two models
agree for a metallicity of $0.1~Z_\odot$. Both simulations use a
25~$\hMpc$ box and $2\times 512^3$ particles. At very high redshift the
metallicity is low and the total SFR is smaller for \emph{SFTHRESZ}
because it has a higher threshold density at this point. Below $z=6$
the situation is reversed, but the difference between the SFHs is
very small, suggesting that the cosmic SFR is dominated by galaxies
that are able to regulate their SFRs.}  
\label{fig:sfthreshold}
\end{figure}

\citet{Schaye2004SF} argued that there is a critical density for the
formation of a cold, interstellar gas phase and that the transition
from the warm to the cold gas phase triggers gravitational instability
on a wide range of length scales. Gas with densities below the
threshold is kept warm ($T\sim 
10^4\,{\rm K}$) and stable by the presence of a UV background. The
predicted critical gas surface density $\Sigma_{\rm g}\sim
3-10~\Msun\,\pc^{-2}$, which agrees well with SF thresholds inferred
from H$\alpha$ observations of nearby galaxies, corresponds to $n_{\rm H}\sim
10^{-2}-10^{-1}\,\cm^{-3}$ for a self-gravitating disk at
$10^4\,\K$. Our reference model uses $n_{\rm H}=10^{-1}\,\cm^{-3}$. 
The \citet{Schaye2004SF} model predicts that the critical density for
SF is a weakly decreasing function of metallicity. We have therefore
run a simulation, \emph{SFTHRESZ},
that uses the predicted scaling (equations 19 and 24 of
\citealt{Schaye2004SF}, valid for $Z = 10^{-4} - 10~Z_\odot$),
\begin{equation}
\label{eq:sfthresz}
n_{\rm H}^*(Z)=10^{-1}\,\cm^{-3} \left ({Z \over 0.1 Z_\odot}\right
)^{-0.64},
\end{equation}
where $Z$ is the gas metallicity and
we used $Z_\odot = 0.02$ for consistency with \citet{Schaye2004SF}. If the 
metallicity is zero then we set $n_{\rm H}^*=10~\cm^{-3}$. 

Fig.~\ref{fig:sfthreshold} compares models \emph{SFTHRESZ} and
\emph{REF}. At very high redshift the metallicity is low and the
threshold density is higher than in the reference
model. This results in a decrease in the SFR that drops rapidly from
0.3~dex at $z=10$ to zero by $z=6$. For $z<6$ the SFR is slightly
higher than in the REF model, which indicates that the metallicity of
the star-forming gas is typically higher than 0.1~solar, but the
effect is marginal. Apparently, after a brief period in which the
SFR is dominated by haloes that are just resolved and therefore just
starting to form stars, the 
predicted SFRs become insensitive to the SF threshold. This suggests that
the galaxies are able to regulate their SFRs. We will provide more
evidence for this below. 

\subsubsection{The Kennicutt-Schmidt star formation law} 
\label{sec:sflaw}

\begin{figure}
\resizebox{\colwidth}{!}{\includegraphics{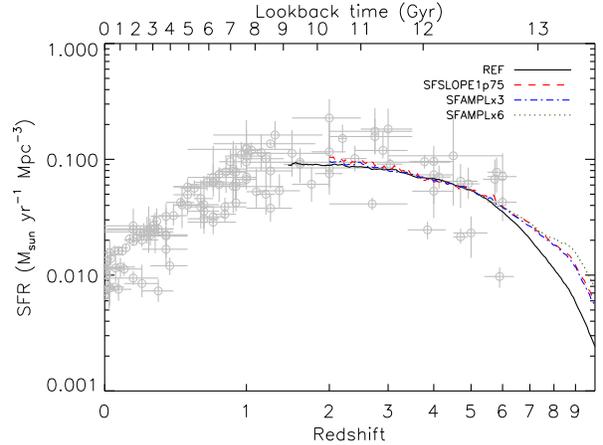}}
\caption{As Fig.~\protect\ref{fig:cosmology},
but comparing the SFHs for models with varying Kennicutt-Schmidt SF
laws. Model \emph{SFSLOPE1p75} (red, dashed) assumes a power-law slope
$n=1.75$ whereas the other models use our fiducial value
$n=1.4$. For models \emph{SFAMPLx3} (blue, dot-dashed) and
\emph{SFAMPLx6} (olive, dotted) the amplitude of the SF law has been
multiplied by factors of 3 and 6, respectively. All simulations use a
25~$\hMpc$ box and $2\times 512^3$ particles. At very high redshift,
when the SFR in the simulations is dominated by poorly resolved haloes, a
more efficient SF law yields a higher SFR. After this initial phase
the SFH is insensitive to the assumed SF law, which suggests that it
is dominated by galaxies that are able to regulate their SFRs.}   
\label{fig:sflaw}
\end{figure}

As discussed in Sec.~\ref{sec:Ref_SF}, gas on the effective EOS is
allowed to form stars at a pressure-dependent rate that reproduces the
observed Kennicutt-Schmidt law \citep{Kennicutt1998Law},
$\dot{\Sigma}_\ast = A (\Sigma_{\rm g}/1~\Msun\,\pc^{-2})^n$, with
$A=1.515\times10^{-4}~\Msolyrkpcsq$ and $n=1.4$. The normalization ($A$)
and slope ($n$) are constrained by
observations, but remain controversial
\citep[e.g.][]{Blanc2009KSlaw}. To develop an understanding of the
physical role of the SF law, we have carried out one run with a
different slope and two with different amplitudes.

Fig.~\ref{fig:sflaw} compares a run with $n=1.75$ (model
\emph{SFSLOPE1p75}; red, dashed) with our 
reference model, which uses $n=1.4$. The SF laws are in both cases
normalized at $\Sigma_{\rm g} = 1~\Msun\,\pc^{-2}$, which is below the
threshold and hence implies that the SFR is higher for all densities
in the run with the steeper slope SF law. For $z>6$
the cosmic SFR is indeed 
higher in the run with $n=1.75$. This is expected, because at these
high redshifts the SFR is dominated by haloes that are just resolved
and therefore just starting to form stars. These galaxies have not yet
had time to become self-regulating and their SFRs are inversely
proportional to the gas consumption time-scales implied by the SF
law. 

Below $z=6$, however, 
the SFRs in the two runs are nearly indistinguishable. This strongly
suggests that the
galaxies are regulating their SFRs such that they produce the
same amount of stars, and thus the same amount of SN energy, 
irrespective of the gas consumption time scale. If a galaxy of a given
halo mass, and hence with a fixed accretion rate, injects too little
SN energy for a galactic outflow to balance the accretion rate, then
the gas fraction, and hence the SFR, will increase. If, on the other
hand, the SN rate is higher than required to balance the infall, then the
gas fraction, and thus the SFR, will decrease. We 
thus expect that when the SF efficiency is changed, the galaxies will
adjust their gas fractions so as to keep their SFR fixed. 
In Haas et al.\ (in preparation) we show that this is indeed what 
happens.  

Finally, Fig.~\ref{fig:sflaw} shows that models in which the
amplitude of the SF law is multiplied by factors of three
(\emph{SFAMPLx3}; blue, dot-dashed) and six (\emph{SFAMPLx6}; olive,
dotted),
respectively, show the same behavior. Initially, the SFR
increases with $A$, but the SFR then quickly asymptotes to a fixed
SFH. Observe that the two runs with higher amplitudes
converge to a common evolution before the reference model joins
them. This is because galaxies can regulate their SF more quickly if
the SF efficiency is higher. Apparently, the cosmic SFR in the reference
model only becomes dominated by self-regulated galaxies by $z=6$. Note
that higher resolution simulations may well find that self-regulation
dominates already at higher redshifts because they can resolve SF in
the progenitors of our lowest mass galaxies.

\subsection{Intermediate mass stars}
\label{sec:chemo}

\begin{figure}
\resizebox{\colwidth}{!}{\includegraphics{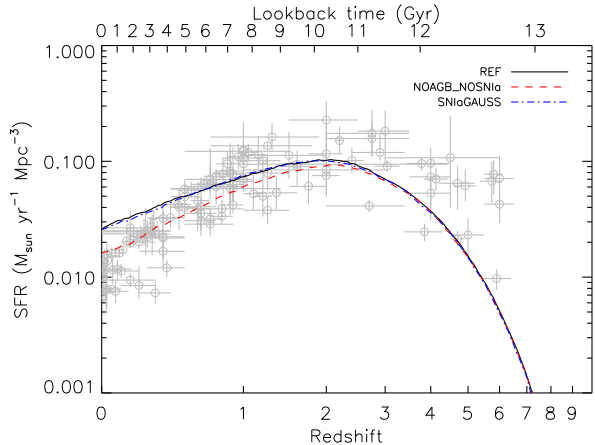}}
\caption{As Fig.~\protect\ref{fig:cosmology},
but comparing the SFHs for models that vary in their treatments of
intermediate mass stars. In model \emph{NOAGB\_NOSNIa} (red, dashed)
mass loss by intermediate mass stars and SN Type Ia has been turned
off. While model \emph{SNIaGAUSS} (blue, dot-dashed) includes these
processes, it assumes 
a Gaussian time-delay function for Type Ia SNe instead of the
e-folding model used in the other simulations. All simulations use a
100~$\hMpc$ box and $2\times 512^3$ particles. While the SNIa time
delay function is unimportant, mass loss by AGB stars provides fresh
fuel for SF and releases metals back into the ISM, thereby
boosting the SFR at late times.}  
\label{fig:chemo}
\end{figure}

Previous numerical studies of the cosmic SFH have mostly used the 
instantaneous recycling approximation \citep[but see e.g.][]{Oppenheimer2008Wmom,Crain2009Gimic}, which means that star particles
eject all the products of stellar evolution immediately following their
formation. Moreover, individual elements are typically not
tracked. Instead, each gas element carries only a single metallicity
variable and relative abundances are assumed to be solar. Furthermore,
such simulations neglect mass loss, i.e., star particles change the
metallicity of their neighbors, but not their masses. As discussed in
section~\ref{sec:Ref_chemo}, we follow the timed release of 11
elements by intermediate mass stars (SNIa and AGB stars) and massive
stars. 

To check the importance of intermediate mass stars, which eject
much of their mass hundreds of millions to billions of years after
their formation and which are responsible for most of the mass lost by
stellar populations, we have run two \emph{L100N512} simulations. In simulation
\emph{NOAGB\_NOSNIa} we do not allow intermediate mass stars to release
mass, leaving massive stars, which evolve on timescales of $\la
10^7\,\yr$, as the only mechanism for releasing metals. To assess the
impact of our choice of the distribution of SNIa progenitor lifetimes,
we ran a simulation (\emph{SNIaGAUSS}) that uses a Gaussian rather than
an e-folding time delay function. This was motivated by the high redshift
observations of \citet{Dahlen2004Rates}, 
which show a marked decline in the SNIa rate beyond $z = 1$.  The
parameters of the delay model are
$\sigma=0.66\,{\rm Gyr}$ and $\tau=3.3\,{\rm Gyr}$
\citep[see][]{Wiersma2009Chemo} .  

Fig.~\ref{fig:chemo} shows that the shape of the SNIa delay function
does not have a significant effect on the predicted SFH. Turning off
both mass loss by SNIa and AGB stars results, however, in a strong
reduction of the SFR at late times. While the reduction factor is
still very small at $z=2$, it increases steadily thereafter to about
0.21~dex at $z=0$, which corresponds to a 40~percent reduction. Given
that the SNIa delay function does not matter, the difference must come
mostly from mass loss by AGB stars. It is not important before $z=2$
because there has not been sufficient time for a substantial fraction
of the stars to reach the AGB phase. Note, however, that higher
resolution simulations will predict higher SFRs at high redshift and
may therefore find that AGB mass loss becomes important earlier. 

Mass loss by AGB stars provides fresh fuel
for SF and releases metals that were locked up in stars. This reduces
the sharpness of the drop in 
the SFR with time, worsening the agreement with
observations. Simulations that ignore this process, will overestimate
the steepness of the drop following the peak in the cosmic SFR.

\subsection{The stellar initial mass function}
\label{sec:imf}

\subsubsection{A Salpeter IMF}
\label{sec:imfsalp}

\begin{figure}
\resizebox{\colwidth}{!}{\includegraphics{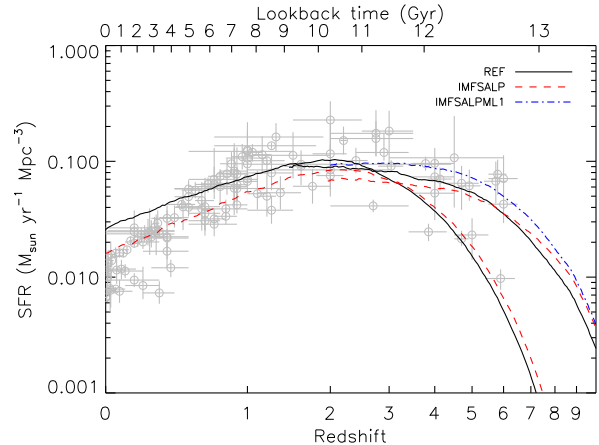}}
\caption{As Fig.~\protect\ref{fig:cosmology},
but comparing the SFHs for models assuming a Salpeter IMF
(\emph{IMFSALP}; red, dashed) to the reference model (black, solid),
which assumes a Chabrier IMF. The amplitude of the SF law is taken
from observations and has therefore been rescaled to the assumed
IMF. It is a factor of 1.65 higher for the Salpeter IMF. Model
\emph{IMFSALPML1} assumes a Salpeter IMF and uses a wind mass
loading factor that is a factor of 1.65 smaller (i.e.\ $\eta=2/1.65$)
than that used in the other models, which accounts for the change in
the number of SNe per unit stellar mass. Note that the observed data
points assume a Chabrier IMF. They need to be shifted upwards by a
factor 1.65 (0.22~dex) to compare with models assuming a Salpeter IMF.
Initially the SFR scales with the amplitude of the SF law and a
Salpeter IMF produces a higher SFR. Later on the SFR is smaller for a
Salpeter IMF because of the decreased importance of metal-line cooling
(because less metals are produced and a greater fraction are locked up
in stars) and stellar mass loss. However, the smaller number of SNe per unit
stellar mass more than compensates for this effect, at least for $z>2$.}  
\label{fig:imfsalp}
\end{figure}

Our reference model assumes a Chabrier IMF, but much of the
literature uses a Salpeter IMF. The two IMFs have similar shapes above
$1~\Msun$, but while the Salpeter IMF is a pure power-law, the
Chabrier IMF includes a lognormal decrease at the low mass end which
results in a much lower stellar mass-to-light ratio. Because most
of the ejected metal mass and all of the energy from core collapse SNe
is produced by
massive stars, the Salpeter IMF is less efficient in enriching
the gas and driving outflows per unit stellar mass formed.

In order to assess the effect of changing the IMF we ran
a simulation employing a Salpeter IMF
(\emph{IMFSALP}), using the same range of stellar masses as we used in
the reference model (i.e.\ $0.1-100~\Msun$). This simulation used
Kennicutt's original normalization for the
amplitude of the SF law ($A =2.5\times10^{-4}~\Msolyrkpcsq$;
\citealt{Kennicutt1998Law}), as he 
assumed the same IMF. Recall that this value is a factor 1.65
greater than the amplitude assumed in \emph{REF} (see
Section~\ref{sec:Ref_SF}).

Fig.~\ref{fig:imfsalp} compares the SFHs predicted by the two
IMFs. Initially the Salpeter IMF gives a slightly higher SFR because
of the higher SF efficiency implied by the change in the
SF law (see also section~\ref{sec:sflaw}). However, after a short
initial phase the SFR falls below that of the reference model. By
$z=0$ the difference has increased to 0.2~dex for \emph{L100N512}. This behavior
can be explained by the fact that a Salpeter IMF produces less
metals and returns less mass (and thus releases less metals that
were locked up in stars) per unit stellar mass formed. Hence,
metal-line cooling is less efficient for a Salpeter IMF. Indeed,
the predicted SFRs fall in between those for the reference model and
the run without metal-line cooling
(c.f.\ Fig.~\ref{fig:fbenergy_zcool0}). 

However, a Salpeter IMF does not only produce less metal mass per
unit stellar mass, it also produces fewer SNe. Assuming that the total
energy in SNe scales as the total number of ionizing photons, the
difference is a factor of 1.65 (Section~\ref{sec:Ref_SF}). Thus, model
\emph{IMFSALP} uses 66~percent of the SN energy to drive winds,
whereas \emph{REF} used only 40~percent. For consistency, we
therefore ran another 
\emph{L025N512} simulation, model \emph{IMFSALPML1}, that is identical to
\emph{IMFSALP}, except that the wind 
mass loading factor was reduced by a factor 1.65 to $\eta =
1.2$. Fig.~\ref{fig:imfsalp} shows that, as expected, 
this run yields a higher SFR than \emph{IMFSALP}, although the two
converge for $z>9$ where there has not been sufficient time for the
simulated galaxies to regulate their SF. In fact, with this change,
a Salpeter IMF yields a higher SFR than a Chabrier IMF. 

\subsubsection{A top-heavy IMF at high pressures}
\label{sec:dblimf}

\begin{figure*}
\resizebox{\colwidth}{!}{\includegraphics{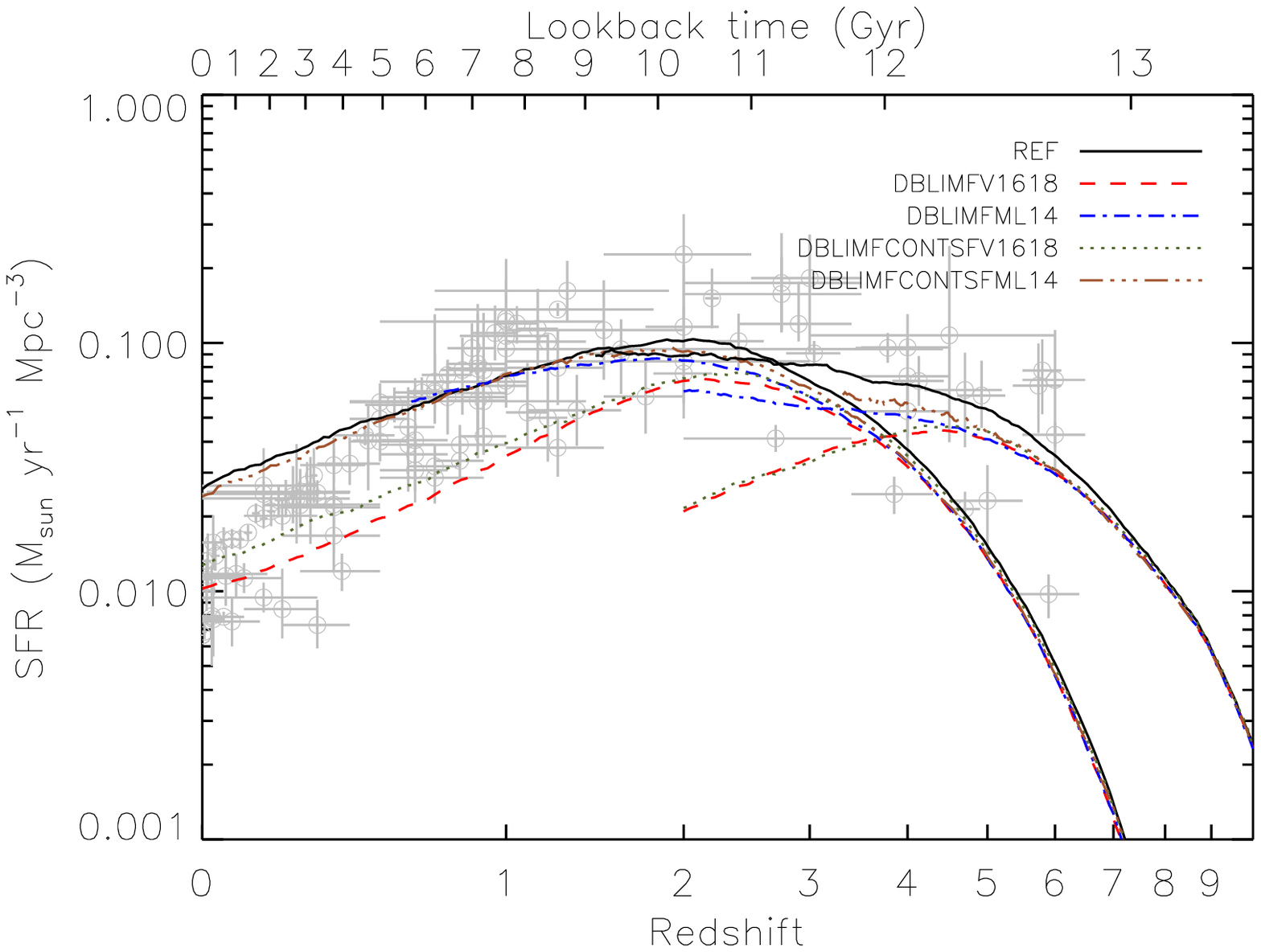}}
\resizebox{\colwidth}{!}{\includegraphics{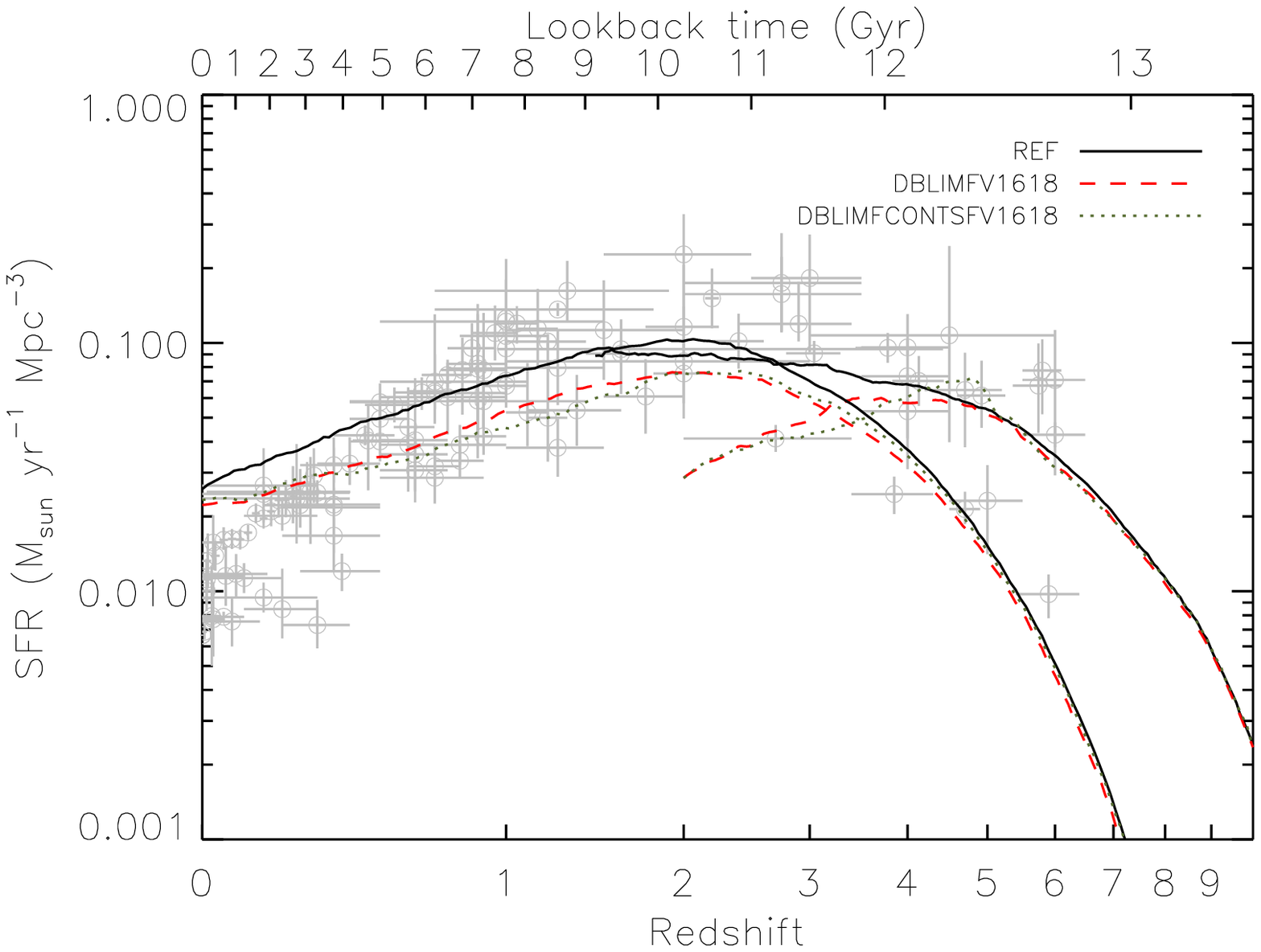}}
\caption{As Fig.~\protect\ref{fig:cosmology}, but comparing the SFHs
  for models with a top-heavy IMF in starbursts. The transition to a
  top-heavy IMF ($dN/dM\propto M^{-1}$ compared to $\propto M^{-2.3}$
  for Chabrier) happens suddenly at a pressure $P/k = 2\times
  10^6\,\cm^{-3}\,\K$. The top-heavy IMF produces 7.3 times more core
  collapse SNe per unit stellar mass formed. Models \emph{ML14} and
  \emph{V1618} use this extra energy to increase the wind mass loading
and velocity, respectively. Models \emph{DBLIMFCONTSF} assume a continuous
SF law, whereas models \emph{DBLIMF} assume the rate of formation of
massive stars to be continuous. The left panel shows actual SFRs,
whereas the SFRs have been rescaled to the ones that would be inferred
under the assumption of a Chabrier IMF in the right panel. Comparisons
to the observed data points, which assumed a 
Chabrier IMF, are only self-consistent for the right panel. The models start
to differ when some 
galaxies have become sufficiently massive to form a fraction of their
stars with a top-heavy IMF. Models with a top-heavy IMF form less
stars, which indicates that the relative increase in the SN rate is
more important than the increase in the metal production
rate. Whether the SF law is continuous or not is unimportant. Using
the extra SN energy to increase the wind mass loading is 
less effective than increasing the wind velocity. A top-heavy IMF in
starbursts reduces the SFR in massive galaxies, but the effect on the
formation rate of massive stars is much less strong.}  
\label{fig:dblimf}
\end{figure*}

Observational determinations of the IMF are extremely difficult. In
particular, extragalactic observations are usually only sensitive to
the light emitted by massive stars, either directly or indirectly via
dust grains. While the IMF is usually assumed to
be universal, it is expected to be top-heavy (or bottom-light) at very
high redshift and low metallicity \citep[e.g.][]{Larson1998TopHeavy}
and both observations and theory 
suggest that it is top-heavy in extreme environments like
the galactic center and starburst galaxies
\citep[e.g.][]{Padoan1997TopHeavy,Baugh2005TopHeavy,Klessen2007TopHeavy,Maness2007TopHeavy,Dabringhausen2009TopHeavy,Bartko2009GCimf}.

We have performed a series of runs to investigate the possible effects
of an IMF that is top-heavy at high pressures, as may be the
case in starbursts and in galactic nuclei. For simplicity, we assume
the IMF switches suddenly from Chabrier to the top-heavy power-law
proposed by \citet{Baugh2005TopHeavy}, $dN/dM\propto M^{-1}$ (as
compared to $\propto M^{-2.3}$ for the high-mass tail of the Chabrier
IMF). The transition is assumed to take place at the
pressure $P/k=2.0\times 10^6\,\cm^{-3}\,\K$ (evaluated at the resolution limit
of the simulations), which was chosen because for this
value $\sim 10^{-1}$ of the stellar mass in our simulations forms at
higher pressures. This ensures that the top-heavy IMF is important,
but not dominant. Of course, a discontinuous dependence on pressure is
not physical, but it is simple and serves to illustrate the
qualitative effects of a top-heavy IMF in starbursts. 

Assuming that the SN energy scales with the
emissivity in ionising photons, our top-heavy IMF yields 7.3 times
more SN energy per unit stellar mass formed. We have therefore
increased the energy injected into the galactic wind by the same
factor for star particles born out of high-pressure gas. In terms of
our kinetic prescription for winds, we can either increase the mass
loading or the wind velocity. We have tried both. Models \emph{ML14}
use a 7.3 times larger mass loading, while models \emph{V1618} use a
$\sqrt{7.3}$ times higher wind velocity. 

If the actual SF law were continuous with pressure, then a sudden
change in the IMF would 
imply a sudden change in the rate of formation of massive stars, which
would manifest itself as a discontinuity in the apparent SF law inferred from
observations under the assumption of a universal IMF. However, the
observed SF law appears to be a continuous power-law (though
\citealt{Krumholz2009SFlaw} suggest that there may
be a kink at $\Sigma_g\sim 10^2\,\Msun\,\pc^{-2}$,
which corresponds roughly to the pressure (see equations 20 and 21 of
\citealt{Schaye2004SF}) at
which we switch IMFs). We therefore tried two possibilities: models
\emph{DBLIMF} (\emph{DBLIMFCONTSF}) assume a continuous
(discontinuous) rate of formation of massive stars, but a
discontinuous (continuous) SF law. 

The left panel of Fig.~\ref{fig:dblimf} compares the SFHs of all four
models for each of 
the two box sizes. Note that models \emph{DBLIMFML14\_L100N512} and
\emph{DBLIMFCONTSFML14\_L025N512} were stopped earlier than the
other runs. The SFHs agree at early times, when SF is confined to
low-mass haloes for which the gas pressure remains low. The models in
which the extra SN energy was used to increase the wind mass loading
predict SFHs that are similar to that of the reference
model. This could mean that increasing the wind mass loading does not
strongly boost the efficiency of SN feedback at high gas
pressures. This agrees well with the results of
\cite{DallaVecchia2008Winds}, who found that at high pressures the
kinetic feedback becomes inefficient due to gas drag and that the
pressure above which this occurs increases with the wind
velocity. Indeed, as can be seen from Fig.~\ref{fig:dblimf}, the
models in which the wind velocity is increased 
for the top-heavy IMF do show a strong reduction in the SFR. 

However, a top-heavy IMF not only yields more SN energy, but also
more metal mass per unit stellar mass formed. The associated increase
in the metal-line cooling rates will boost the SFRs (see
Section~\ref{sec:zcool0}). The relatively small difference between the
\emph{ML14} and \emph{REF} model could therefore also mean that the
increased wind mass loading compensates for the higher cooling rates. 

The differences between \emph{DBLIMF} and the corresponding
\emph{DBLIMFCONTSF} runs are small. This was to be expected, as we already
showed in section~\ref{sec:sflaw} that the SFHs are insensitive to the SF law
because galaxies regulate their gas fractions so as to produce the
same amount of SN energy, irrespective of the assumed SF
efficiency. 

To compare with the observed data points, which were derived from
observations of massive stars under the assumption of a Chabrier IMF,
we have to multiply the rate of SF in the top-heavy mode by a factor
7.3. The right panel of Fig.~\ref{fig:dblimf} shows that doing so reduces the
drop at late times, i.e.\ the SFR inferred under the assumption of a
universal IMF falls off less steeply than the actual SFR. Observe that
the differences between models \emph{DBLIMF} and \emph{DBLIMFCONTSF}
are also reduced, particularly for $z<1$. This supports our proposal
that because galaxies 
regulate their SF, they inject a fixed amount of SN energy for a
given halo mass. 

Finally, we note that the agreement between the 25 and 100~$\hMpc$
boxes is much poorer for the models with a top-heavy IMF in starbursts
than it was for the other models. This reflects our choice to make the
IMF a function of the pressure, which is 1-1 related to the gas
density in our simulations since star-forming gas follows a polytropic
EOS. Increasing the resolution decreases the mass above which haloes
contain enough particles to sample the high-density tail of the gas
distribution. Hence, lower halo masses will be able to form some
fraction of their 
stars with a top-heavy IMF and thus suppress subsequent SF. Clearly,
using prescriptions for feedback that are functions of density or
pressure will make the results more prone to resolution effects.

We conclude that although our toy models are too simple and sensitive
to resolution, it is clear that a top-heavy IMF in starbursts can
serve to suppress SF 
in high mass haloes. This can result in a steeper drop in the SFR at
late times, as suggested by observations (although the
effect is less strong when only
the rate of formation of massive stars is considered). The stronger
suppression in massive haloes can also shift the peak in the SFH to
higher redshifts.
 
\subsection{SN-driven winds}
\label{sec:sn}

It is well known that simulations without galactic winds suffer from a
severe overcooling problem: the SFR greatly exceeds the observational
constraints and is usually only limited by numerical resolution
\citep[e.g.][]{Balogh2001Overcooling}. As Fig.~\ref{fig:fbenergy_zcool0} shows,
except at very high 
redshift when the SFR in the real Universe is expected to be dominated
by haloes that are below the resolution limit of our simulations, the
runs without SN feedback do indeed produce far too 
many stars. Moreover, comparison of the SFHs for models \emph{REF} and
\emph{WML4} in Fig.~\ref{fig:cosmology} shows that doubling the
SN energy that is injected, in this case by doubling the wind mass
loading, further reduces the SFR.

In sections~\ref{sec:sflaw} and \ref{sec:dblimf} we showed that the SFH is
insensitive to the assumed SF law. We concluded from this that SF in
galaxies is self-regulating: the SFR adjusts such that outflows driven
by feedback from massive stars balance the infall driven by gas
accretion onto haloes and radiative cooling. If the SF law is changed, then
galaxies simply adjust their gas
fractions in order to inject the same amount of SN energy into haloes
of a given mass. If galaxies do indeed regulate their SFRs in this
manner, then we would expect the rate of
energy injection into the winds to remain constant if the fraction of
the SN energy that is injected is varied. In other words, the SFR
should be inversely proportional to this fraction. We can test this by
comparing model \emph{REF} to simulation \emph{WML4}, which injects
twice as much SN 
energy per unit stellar mass. Fig.~\ref{fig:cosmology} shows that while
the SFR is indeed lower for \emph{WML4}, the difference is always
smaller than 0.25~dex, whereas we would have expected 0.30~dex
(i.e.\ a factor of two) at late times, when the SFR is dominated by
galaxies that have had enough time to become self-regulating. 

There are, however, good reasons why we would not expect the SFR to be
exactly inversely proportional to the efficiency of the SN
feedback. First, a change in the SFR does not only change the rate of
energy injection into winds, it also changes the rate of metal
injection and the rate at which mass loss from AGB stars supplies the
galaxy with fresh fuel for SF. Both of these effects would, however,
tend to lower the amount of SN energy that is needed for
self-regulation, which means that we would expect the SFR to decrease
faster than linear with the SN efficiency. This is opposite to what is
actually happening. Second, a reduction in the SFR implies a reduction in the
stellar mass and hence in the gravitational force. Again, this would
imply that less SN energy is required for self-regulation, which would lead
to an even larger drop in the SFR, contrary to what the simulations
predict. 

The reason why the SFR varies more slowly with the SN efficiency than
the inverse proportionality we would naively expect, is likely that
SN winds are not effective in high mass galaxies, at least when
kinetic feedback is used with a velocity of $600~\kms$. If the
feedback is inefficient, then we would not expect the galaxies to be
able to self-regulate. Indeed, we will show in Haas et al.\ (in
preparation) that, at
$z=2$, the SFR in \emph{WML4} is in fact 
half that of \emph{REF} for haloes with total mass less than
$10^{11}\,\Msun$ and that the SN feedback becomes inefficient
for higher halo masses.

\subsubsection{Varying the parameters at constant wind energy}
\label{sec:winds_ce}

As discussed in section~\ref{sec:Ref_sn}, we inject the energy from SNe
in kinetic form. Newly formed star particles kick their gaseous
neighbours with a constant velocity $v_{\rm w}$ in a random
direction. On average, the mass kicked is $\eta$ times the
mass of the star particle. While the product $\eta v_{\rm w}^2$ determines the
energy of the winds and is therefore constrained by the energy
available from SNe, it is not clear a priori what values should be
chosen for the individual parameters. Note that they cannot be 
taken directly from observations because the parameter values refer to the
properties of the wind at the inter-particle distance (neighbours of new
stars) which varies and will typically not agree with the scales
relevant for the observations. The observational constraints are usually
inferred from  the velocity offset and column densities of blueshifted
absorption lines, but it is unclear at what distance from 
the source the absorption occurs
\citep[e.g.][]{Veilleux2005windsreview}. Moreover, the absorption
lines probe only the cold part of the outflow. The constraints on the
velocity and mass loading of the hot wind are also very poor. 
 
Given the lack of observational constraints, one would hope that the
results are insensitive to the 
amount of mass that a fixed amount of SN energy is distributed over. 
\citet{DallaVecchia2008Winds} showed that this is indeed the
case for the SFRs provided the galaxies are well resolved and the wind
velocity exceeds a critical value that increases with the pressure of
the ISM and thus also with halo mass. If, however, the wind velocity
is too low, then the wind particles are immediately stopped by drag forces
and never leave the ISM. If the disks are
unresolved, then the 
hydrodynamic drag is underestimated and for sufficiently low
resolutions all particles that are kicked are
able to escape the ISM.

The runs with a top-heavy IMF in starbursts
(Fig.~\ref{fig:dblimf}) show that SF is much more efficiently
suppressed if the extra SN energy (relative to a Chabrier IMF) is used
to increase the wind velocity $v_{\rm
  w}$ than if it is used to increase the mass loading factor
$\eta$. Since in these models 
the feedback energy is only boosted in high-pressure gas, this suggests that
the wind velocity of $600~\kms$ that was used in the reference model
is insufficient at the high pressures that we required for the IMF to become
top-heavy. This implies that our default prescription for SN feedback
is inefficient in high mass galaxies, which could account for the fact
that the SFR drops off less rapidly at late times than is observed.

\begin{figure}
\resizebox{\colwidth}{!}{\includegraphics{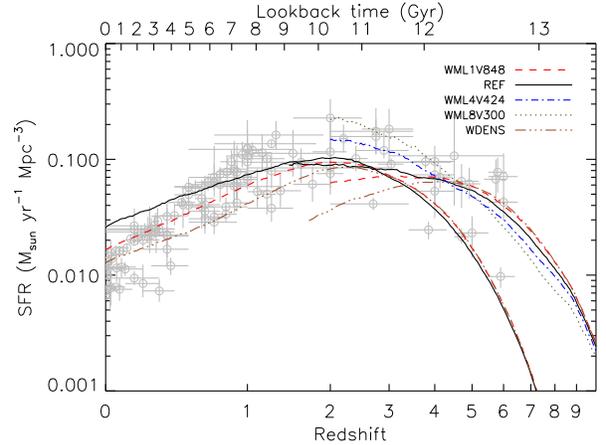}}
\caption{As Fig.~\protect\ref{fig:cosmology},
but comparing the SFHs for models that all inject the same amount of
SN energy per unit stellar mass (i.e.\ $\eta v_{\rm w}^2$ is constant),
but that assume different wind velocities. Models \emph{WMLxVyyy}
assume a wind mass loading $\eta=x$ and velocity $v_{\rm w}=yyy~\kms$. Model
\emph{WDENS} assumes that the wind velocity scales with the local
sound speed, which implies $v_{\rm w}\propto \rho^{1/6}$ for our EOS,
normalized to the value used in the reference model at the SF
threshold. At high redshift, when the SFR is dominated by poorly
resolved, low-mass haloes, the SN feedback is more efficient if it is 
distributed over more mass. However, at low redshift the situation is
reversed. This can be explained if the feedback becomes
inefficient when the velocity falls below a critical value that
increases with galaxy mass. Scaling the wind parameters with local
properties, such as the density for model \emph{WDENS}, can help 
to keep the feedback efficient, but it also makes the results sensitive
to the resolution.}  
\label{fig:winds_ce}
\end{figure}

To further investigate the dependence on the two individual wind
parameters, we have run a series of simulations that all inject the same
amount of SN energy per unit stellar mass as the reference model, but
assume mass loading factors that 
differ by factors of two, ranging from 1 to 8 in the 25~$\hMpc$ box
($v_{\rm w}\propto \eta^{-1/2}$ varies from 848 to $300~\kms$) and
from 1 to 2 in the 100~$\hMpc$ box ($v_{\rm w}$ varies from 848 to $600~\kms$). 
Fig.~\ref{fig:winds_ce} compares the SFHs
of these runs. Clearly, the results are not just determined by the
total energy injected into the wind, which is identical for all the
runs. At high redshifts the SFHs are similar, 
although the feedback is slightly 
more efficient for higher values of $\eta$. However, at late times 
the different SFHs start
to diverge, with higher wind velocities suppressing the SF more
strongly. 

These results are consistent with the conclusions of
\citet{DallaVecchia2008Winds}. As the universe evolves, stars typically
form in more massive haloes and the minimum wind velocity for which
the feedback remains 
effective thus increases. Hence, the redshift for which the feedback
becomes inefficient decreases with increasing wind velocity. At early
times, when many stars form in haloes that are poorly resolved, higher
mass loading factors are more efficient because all particles that are
kicked from poorly resolved haloes are able to escape the ISM. 

The SFH, including the redshift at which it peaks, is clearly
sensitive to the poorly constrained parameters $\eta$ and $v_{\rm
  w}$. The same is likely to be 
true for other types of subgrid prescriptions than kinetic
feedback. Thus, unless one varies the parameters of the wind model,
which is unfortunately not always done in the literature, 
one risks overinterpreting the results.

If the main goal were to reproduce the observed SFH and if one were
willing to accept the lack of ``ab initio predictive
power'' displayed by Fig.~\ref{fig:winds_ce}, then one could choose to
take an approach similar to that of 
semi-analytic models and take advantage of the sensitivity of
the results to the wind parameters. By varying the parameters with
halo mass or with the physical properties of the star-forming gas, one
could match a wide range of SFHs. 

While we have not tried to tune the SFH, we
have investigated a toy model that uses the same amount of SN energy as
the reference model, but in which the wind velocity scales with the
local effective sound speed, $c_{\rm s,eff}$, as might be the case for thermally
driven winds. If 
it is indeed hydrodynamic drag that stalls low-velocity winds, then
this scaling could keep the winds efficient at all pressures. We
implement this model, which we term \emph{WDENS}, by making the wind
parameters functions of the 
density of the gas from which the star particle formed:
\begin{eqnarray}
v_{{\rm w}} &=& v_{\rm w}^\ast \left (\frac{\nH}{\nHs}\right )^{1/6}, \\
\eta &=& \eta_\ast \left (\frac{v_{\rm w}}{v_{\rm w}^\ast}\right
)^{-2} = \eta_\ast \left (\frac{\nH}{\nHs}\right )^{-1/3},
\end{eqnarray}
which implies $v_w \propto c_{\rm s,eff}$ since star-forming gas follows the
effective EOS $P = \rho_{\rm g} c_{\rm s,eff}^2 \propto
\rho_{\rm g}^{4/3}$. We set $v_{\rm w}^\ast=600~\kms$ and $\eta_\ast=2$,
so that the values of the wind parameters agree with those of
the reference model for stars formed at the density threshold $\nHs$,
while the wind velocity is greater (and the mass loading smaller) at
higher pressures. 

Comparing the 25~$\hMpc$ runs shown in Fig.~\ref{fig:winds_ce}, we see that
\emph{WDENS} predicts a nearly identical SFH as \emph{WML1V848} and
\emph{REF} down to $z=4$, but that it generates much more efficient
winds at later times. The 100~$\hMpc$ shows qualitatively the same
behavior, with \emph{WDENS} predicting significantly lower SFRs below
$z=2$. Because the winds in \emph{WDENS} remain effective for higher
galaxy masses, the drop in the SFR below redshift 2 is steeper
than for the reference model, although it is still less steep than
observed. 

Comparing the two \emph{WDENS} runs, we see that the
agreement between the different box sizes is much worse than for the
reference model. Clearly, making SN feedback a function of the local
gas density increases the sensitivity to numerical resolution as we
already concluded from the models that used a top-heavy IMF at high
densities (see section~\ref{sec:dblimf}). This is probably because the
high density tail of the PDF can only be sampled if the galaxy
contains a sufficient number of particles. One would therefore expect
somewhat better convergence if the wind parameters were a function of the
properties of the dark matter halo rather than the local gas
properties. We will investigate such models in
Section~\ref{sec:wmom}. 

\subsubsection{Hydrodynamically decoupled winds}
\label{sec:hydrodec}

\begin{figure}
\resizebox{\colwidth}{!}{\includegraphics{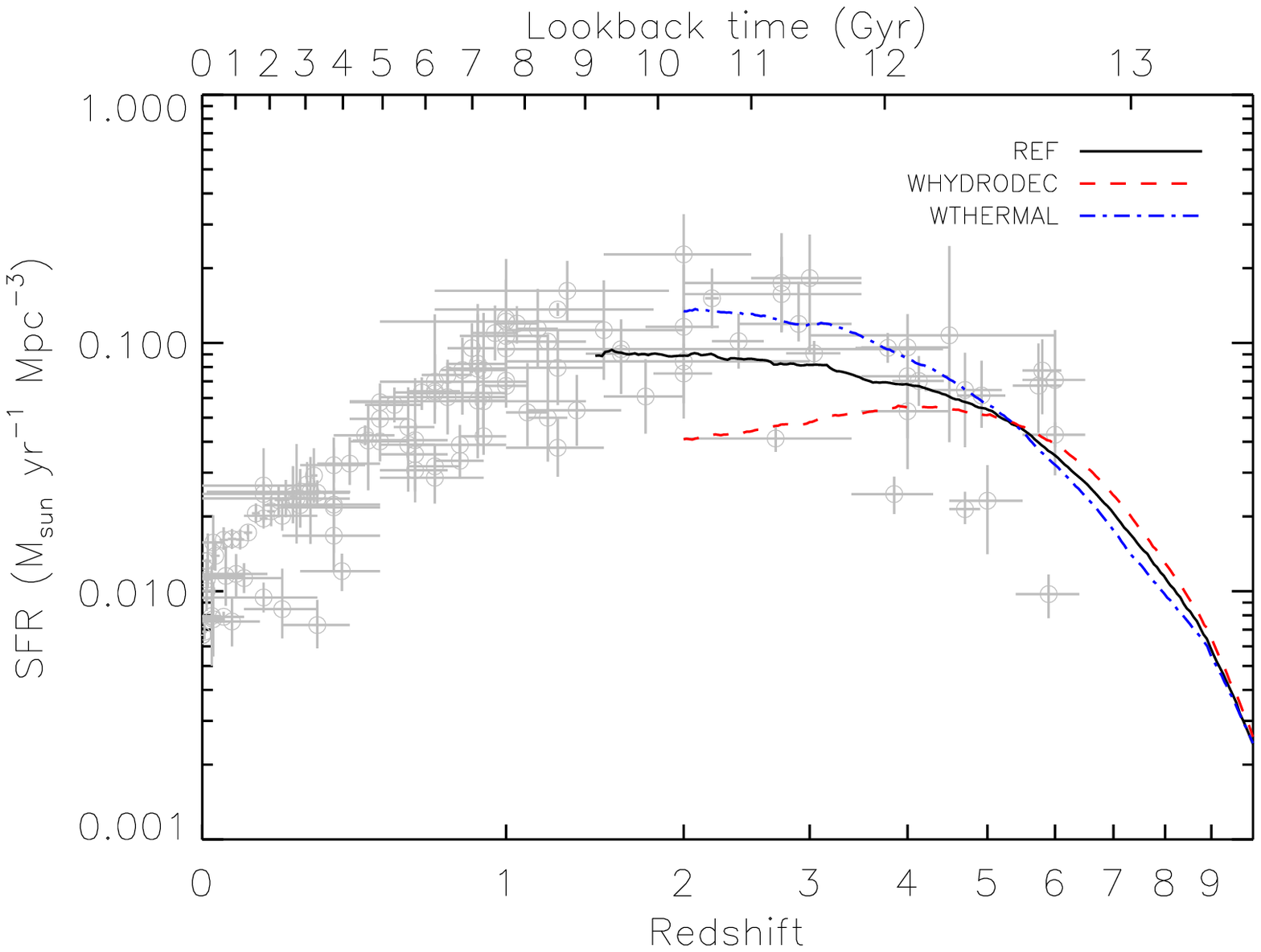}}
\caption{As Fig.~\protect\ref{fig:cosmology}, but comparing the SFHs
  for models that use different implementations of SN feedback. Model
  \emph{WHYDRODEC} is identical to the reference model, except that
  the wind particles are temporarily decoupled from the
  hydrodynamics as in \protect\citet{Springel2003Multiphase}. Model
  \emph{WTHERMAL}, on the other hand, injects the SN energy in thermal
  form following the prescription of
  \protect\citet{DallaVecchia2009Winds}. Hydrodynamically decoupled
  wind particles can freely escape the ISM, but are unable to drag
  other particles along. They are therefore less efficient at high
  redshift, when low-mass galaxies dominate the SFR, but they are much
  more efficient at low redshift, when hydrodynamical drag within the
  high-pressure ISM of massive galaxies is important. Injecting the
  same amount of SN
energy in thermal form increases the efficiency of the feedback in
poorly resolved, low-mass galaxies, but the winds are somewhat less
effective at low redshift, i.e.\ for higher mass galaxies.}  
\label{fig:whydrodec_thermal}
\end{figure}

In recent years, a large fraction of the results from cosmological,
SPH simulations discussed in the literature were obtained from simulations
run with \textsc{gadget2} \citep{Springel2005Gadget2} and employing the
\citet{Springel2003Multiphase} prescription for kinetic SN 
feedback. This 
prescription for galactic winds differs in two respects from
ours. First, the
wind particles are selected stochastically from all the star-forming
(i.e.\ dense) particles in the simulation and are therefore not local
to the star particles as is the case for us.
Second, the wind particles are subsequently decoupled from the hydrodynamics
for 50~Myr (i.e.\ 31~kpc if traveling at $600~\kms$) or until
their density has fallen below 10 per cent of the threshold for SF,
which ensures that they escape the ISM. 

\cite{DallaVecchia2008Winds} investigated the effects of this
decoupling in detail and found them to be dramatic. While decoupled
winds remove fuel for SF, they cannot blow bubbles in the disc,
drive turbulence or create channels in gas with densities typical of
the ISM. Decoupled winds are less efficient at suppressing SF in low
mass galaxies, because they cannot drag gas along. They are, however,
much more efficient for high mass galaxies, because they do not suffer the
large energy losses due to drag in the high-pressure ISM. As the
numerical resolution is decreased, the disc responsible for the drag disappears
and the two prescriptions tend to converge. Hence, the decoupled winds
are less sensitive to resolution, essentially because they only need
to resolve the Jeans scales at the relatively low density for which
the wind particles are recoupled. 

Fig.~\ref{fig:whydrodec_thermal} shows the SFH predicted by a model
in which the wind particles were decoupled from the hydrodynamics in
the manner described in \cite{Springel2003Multiphase}
(\emph{WHYDRODEC}) (note that the wind particles were, however, still
local to the newly formed star particles). Compared with the reference
model, the decoupled winds are less efficient at high redshift,
because the wind particles cannot drag other gas particles out of low-mass
galaxies. We expect this difference to increase for higher resolution
simulations, because they can resolve the discs of such
galaxies better. Decoupled winds are, however, much more effective at lower
redshifts when 
the galaxies dominating the SFR are better resolved and more massive,
leading to large energy losses due to drag in the ISM for the
reference model. Consequently, the peak in the SFR shifts from $z\sim 2$ for
the reference run to $z\approx 4$ for the decoupled winds. 

\subsubsection{Thermal SN feedback}
\label{sec:thermal}

\begin{figure}
\resizebox{\colwidth}{!}{\includegraphics{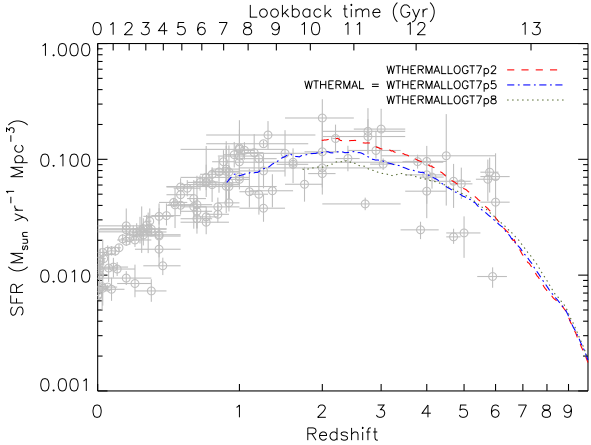}}
\caption{As Fig.~\protect\ref{fig:cosmology}, but comparing the SFHs
  for models that all employ thermal SN feedback using the
  method of \protect\citet{DallaVecchia2009Winds}. All models inject
  40~per cent of the available SN energy per unit stellar mass, but they use
  different temperature jumps for the heated gas. Models
  \emph{WTHERMALLOGTxpy} use temperature jumps $\log_{10}\Delta T = 
  x.y$ (e.g.\ \emph{LOGT7p5} implies $\Delta T = 10^{7.5}\,\K$)
  . Model \emph{WTHERMALLOGT7p5} assumed identical parameters 
  as model \emph{WTHERMAL} shown in
  Fig.~\protect\ref{fig:whydrodec_thermal}. All simulations use a
  12.5~$\hMpc$ box and $2\times 256^3$ particles. Enforcing a greater
  temperature jump, which implies a smaller heating probability, makes
  the feedback less efficient at high redshift, but more efficient at
  low redshift. However, the feedback efficiency is less sensitive to
  the temperature jump than it is to the wind velocity for the case of
  kinetic SN feedback.}
\label{fig:wthermal}
\end{figure}

In the previous sections we found that the results of simulations
employing kinetic SN feedback 
are sensitive to the parameters of the subgrid model, even if the
total energy in the wind is kept constant. It is therefore of interest
to consider alternatives to kinetic feedback. In particular, thermal
feedback would seem a natural choice as it would allow the simulation
itself to determine the properties of the wind such as the mass
loading, velocity and geometry, based on the properties of the
starburst. Unfortunately, 
current simulations of cosmological volumes cannot
resolve the energy-conserving 
phase in the evolution of SN remnants, causing any thermal energy
input to be mostly radiated away before it can be converted into
kinetic form. This is why alternatives such as kinetic feedback have
been developed in the first place.

As discussed in \citet{DallaVecchia2009Winds}, this
overcooling problem is mainly caused by the fact that in naive
implementations of thermal feedback the ratio of the heated mass to
that of the star particle is too large, which means the temperature
jump is too small and hence that the radiative cooling times are too short. The
problem can thus be overcome if the SN energy produced by a
star particle is injected in a sufficiently small amount of mass, such
that its cooling
time becomes long compared to the time scale on which the local gas
density can change in response to the energy injection. 
\citet{DallaVecchia2009Winds} propose a stochastic
method, which generalizes a prescription introduced by
\citet{Kay2003Xraygroups}, that uses the temperature increase of the
heated gas, $\Delta T$, and the fraction of the SN energy that is
injected, $f_{\rm th}$, as parameters. These parameters then determine
the probability for a 
gas element neighbouring a newly formed star particle to be
heated. \citet{DallaVecchia2009Winds} demonstrated that as long
as the temperature increase is greater than a value that depends
weakly on the resolution and the gas density, the factor by which the
feedback suppresses SF is insensitive to the value of the
temperature increase.

We have performed a run, \emph{WTHERMAL}, in the 25~$\hMpc$ box using
the thermal feedback prescription of 
\citet{DallaVecchia2009Winds}, setting $\Delta T = 10^{7.5}\,\K$ and
$f_{\rm th} = 0.4$. The latter value agrees with the one implied by
the product $\eta v_{\rm w}^2$ of the parameters of the kinetic
feedback used in the reference run. Substituting these parameter
values into the equations presented in
\citet{DallaVecchia2009Winds}, we estimate that for our resolution
the thermal feedback 
will be efficient up to at least the density $n_{\rm H}\sim 2\times
10^2\,\cm^{-3}$, which exceeds our SF threshold by more than three
orders of magnitude. Per star particle, the average number of gas
particles that receive SN energy is about 0.54.

Fig.~\ref{fig:whydrodec_thermal} shows that the thermal feedback, when
implemented in this manner, is indeed effective at suppressing the
SFR. At high redshift it is more efficient than the kinetic feedback
used in the reference model, which suggests that it results in higher mass
loading factors for poorly resolved galaxies. For $z<5$ the SFR is
higher than in the reference run, but the difference is always less
than 0.2~dex and remains constant below redshift 3.5. 

Fig.~\ref{fig:wthermal} shows the SFHs for three different models
that all employ thermal feedback using identical amounts of SN energy
($f_{\rm th}=0.4$), but different temperature jumps. From
top-to-bottom at redshift 2, the models use $\Delta T = 10^{7.2}$,
$10^{7.5}$ 
and $10^{7.8}\,\K$, respectively. Hence they differ by factors of two,
which matches the factors of $\sqrt{2}$ difference between the wind
velocities used in Fig.~\ref{fig:winds_ce}. These three
thermal feedback models use $2\times 256^3$ particles in a
12.5~$\hMpc$ box, which means the resolution is identical to that used
in the \emph{L025N512} runs shown in Fig.~\ref{fig:winds_ce}.

Comparing the three models, we see that a
higher temperature increase makes the feedback slightly less efficient
at high redshift ($z>6$), although the effect is
marginal. Interestingly, the small difference at $z>6$ appears to be
be caused by the varying strength of the response to the reheating
associated with reionization (which happens at $z_{\rm r}=9$ in our
simulations). This agrees with the finding of
\citet{Pawlik2009Photowinds} that SN feedback and photo-heating
strengthen each other. Near the end of the simulations the feedback is
more efficient for greater $\Delta T$. The 
redshift at which $\Delta T=10^{7.8}\,\K$ becomes more
efficient than $\Delta T=10^{7.5}\,\K$ is lower than the redshift at
which the latter model becomes more efficient than $\Delta
T=10^{7.2}\,\K$. All of this can be explained if the thermal feedback
becomes inefficient for haloes more massive than some value which
increases with $\Delta T$. This situation parallels that of the
kinetic feedback with $\Delta T$ playing the role of $v_{\rm
  w}$. 

Interestingly, the SFH is much less sensitive to the value of $\Delta T$ than
it is to $v_{\rm w}$ for the case of kinetic SN feedback. The
difference between the models with $\Delta T = 10^{7.2}$ and 
$10^{7.8}\,\K$ is always less than 0.24~dex, whereas the difference
between the models with $v_{\rm w} = 424$ and $848~\kms$ is 0.38~dex
at $z=2$ (see Fig.~\ref{fig:winds_ce}). Moreover, while the SFHs in
the different
kinetic feedback models diverge rapidly, the differences between the
thermal models is nearly constant below $z=3$.

Our findings that the thermal feedback is efficient and that it is less
sensitive to the parameters of the model than is the case for kinetic
feedback are very encouraging. We only used 40~per cent of the SN
energy because we wanted to match the fraction used for the kinetic feedback in
the reference model. 
Higher values can, however, easily be justified for thermal feedback
since we are now simulating radiative losses.

\subsection{``Momentum-driven'' winds}
\label{sec:wmom}

\begin{figure}
\resizebox{\colwidth}{!}{\includegraphics{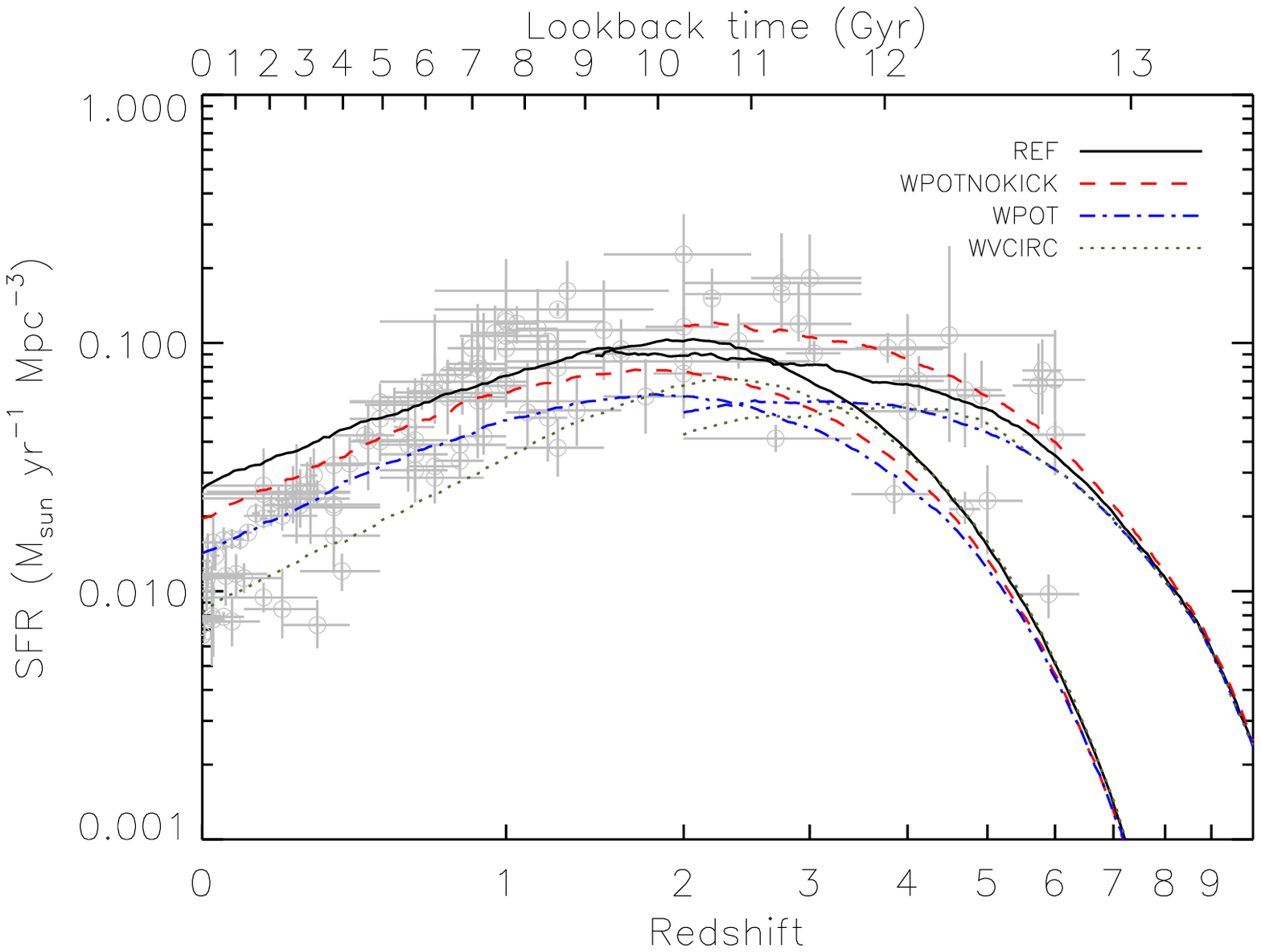}}
\caption{As Fig.~\protect\ref{fig:cosmology},
but comparing the SFHs for different implementations of
``momentum-driven'' winds, following
\protect\citet{Oppenheimer2006Wpot,Oppenheimer2008Wmom}. While these
models were motivated by the 
claim that galactic outflows may be driven by radiation pressure on
dust grains \protect\citep{Murray2005MomWinds}, they do not
actually include radiative 
transfer. Instead, the parameters of the kinetic feedback prescription
are made functions of either the local potential (\emph{WPOT} and
\emph{WPOTNOKICK}) or the mass of the dark matter halo
(\emph{WVCIRC}). The wind parameters scale such that the wind velocity
increases with halo mass, while the mass loading is inversely
proportional to the wind velocity. For massive galaxies the
winds use more energy than is available from SNe. See the text for
additional details. Because the wind velocity increases with halo
mass, the feedback is more efficient than for the reference model at
late times. At high redshift it is also somewhat more efficient thanks
to the higher mass loading factors.}   
\label{fig:wmom}
\end{figure}

We have so far only considered galactic winds driven by SN
feedback. It is, however, also possible that outflows are driven by
the momentum that is deposited by photons from massive stars as they
are absorbed by dust grains \citep{Murray2005MomWinds}. In such a
momentum-driven wind one would expect the mass loading to be inversely
proportional to the wind velocity and, according to the model of
\citet{Murray2005MomWinds}, the terminal wind velocity would be similar
to the velocity dispersion of the host galaxy. \citet{Martin2005Winds}
argued that such scalings are in agreement with estimates of the mass
and velocity in cold clouds as inferred from \NaI\ absorption
lines. \citet{Oppenheimer2006Wpot} implemented several versions of 
momentum-driven winds into cosmological simulations of the chemical
enrichment of the IGM and found good agreement with observations of
\CIV\ absorption.  

Direct simulation of radiation pressure requires radiative transfer, 
which is too costly for cosmological
simulations. \citet{Oppenheimer2006Wpot} therefore chose to use the 
\citet{Springel2003Multiphase} recipe for kinetic feedback, including
the hydrodynamic decoupling discussed in section~\ref{sec:hydrodec}, but to
make the velocity kick, $v_{\rm w}$, and the mass
loading factor, $\eta$, functions of the properties of the galaxy so as to
mimic the scalings expected for momentum-driven winds. Specifically, they
assumed 
\begin{eqnarray}
v_{\rm w} &=& \left (a_1 \sqrt{a_2 f_{\rm L}(Z)-1} + a_3\right )\sigma, \\
\label{eq:wmom_vw}
\eta &=& a_4 / \sigma,
\end{eqnarray}
where $a_1 - a_4$ are free parameters, $\sigma$ is the galaxy velocity
dispersion, and $f_{\rm L}(Z)$ is a function that accounts for the dependence of
the stellar 
luminosity on metallicity and that varies from 1.7 at
$10^{-3}\,Z_\odot$ to unity for solar abundances. The parameter $a_1$
was set to 3 which is the value suggested by
\citet{Murray2005MomWinds}. Parameter $a_2$ was either set to 2 or
varied randomly between 1.05 and 2. Parameter $a_3$ was introduced after
noting that for radiatively driven winds the outflow velocity
increases out to large distances, whereas in the simulations
the gas is not given any more momentum after it is \lq kicked\rq\ out of
the ISM. They tried both $a_3=0$ and $a_3=2$. Finally, parameter
$a_4$ was set to $300~\kms$ in order to roughly match the observed
cosmic SFR at high redshift. The velocity dispersion of the host
galaxy was estimated by taking the local gravitational potential,
$\Phi$, and estimate $\sigma$ using the virial theorem
($\sigma=\sqrt{-\frac{1}{2}\Phi}$).

Later papers by the same authors compared to other types of
observations, but used different parameter values and 
methods. \citet{Oppenheimer2008Wmom} realized that the
gravitational potential is dominated by large-scale structure rather
than by the mass of individual haloes. For this reason, they moved
away from using the 
gravitational potential to estimate wind properties and instead used
friends-of-friends halo catalogues,
generated on-the-fly throughout the simulation, and set $\sigma =
\sqrt{2} v_{\rm c}$ where $v_{\rm c} = \sqrt{GM/R_{\rm vir}}$ is the
halo circular velocity and $R_{\rm vir}$ is the virial radius. 
\citet{Oppenheimer2008Wmom} also
increased $a_1$ from 3 to 4.3, halved the value of $a_4$ to $150~\kms$
and imposed an upper limit on 
the wind velocity corresponding to twice the total SN
energy. Later papers used the same values as
\citet{Oppenheimer2008Wmom} although \citet{Oppenheimer2009OVI} no 
longer imposed a limit on the total wind energy. 

We first implemented the \citet{Oppenheimer2006Wpot} method, which uses the
local potential to estimate the velocity 
dispersion, but without
decoupling the wind particles from the hydrodynamics. We used the
parameter values advocated by 
\citet{Oppenheimer2008Wmom}, although we neglected the
metallicity-dependent square root term, which simplifies
equation~(\ref{eq:wmom_vw}) to $v_{\rm w} = (a_1+a_3)\sigma$. We ran
two versions. While both assumed $a_1=3$ and $a_4=150~\kms$, model \emph{WPOTNOKICK}
used $a_3=0$ whereas model \emph{WPOT} used
$a_3=2$. Fig.~\ref{fig:wmom} shows the SFHs predicted by the models
for both box sizes. \emph{WPOT} gives lower SFRs than
\emph{WPOTNOKICK} which is not surprising since it uses higher wind
velocities. Model \emph{WPOTNOKICK} predicts higher SFRs than the reference
model for the 25~$\hMpc$ box, but the order is reversed for the
100~$\hMpc$ box. This reflects the fact that there is more large-scale
structure in the larger box and that the potential is dominated by the
largest structures in the box. Clearly, this situation is not
desirable. 

We therefore also ran model \emph{WVCIRC} which
estimates $\sigma$ using an on-the-fly halo friends-of-friends halo finder
as in \citet{Oppenheimer2008Wmom}\footnote{Note that \citet{Oppenheimer2008Wmom}
  ran a friends-of-friends halo finder on the baryons using a linking
  length of 0.04 
  and assuming a fixed baryon to dark matter ratio equal to the
  universal value.}. The halo finder was run at times
spaced evenly in $\log a$ with $\Delta a = 0.02a$, where $a$ is the
expansion factor. Haloes were
found based on the distribution of dark matter particles using a
linking length of 0.2 and requiring a minimum of 25 dark matter
particles per halo\footnote{For star particles forming outside of any
  halo we assume the minimum halo mass corresponding to 25 dark matter
particles.}. Baryonic particles were attached to the nearest
dark matter particle.

The predicted SFHs are also shown in
Fig.~\ref{fig:wmom}. Except at very high redshift, the SFR is strongly
reduced compared to the reference model. The difference increases with
time, so that the SFR falls off more rapidly below $z=2$ than for
\emph{REF}, as required
by the observations. As was the case for \emph{WDENS} (see
Section~\ref{sec:sn} and Fig.~\ref{fig:winds_ce}), the increased
efficiency of the feedback at 
late times arises 
because the wind velocity increases with the halo mass, whereas it is
constant for the reference model. Contrary to
\emph{WDENS}, the feedback is also more efficient than that of the
reference model at high redshift. This is a consequence of the high
mass loading used for low-mass haloes (recall that for \emph{WDENS}
the mass loading is never higher than for \emph{REF}). 

The implementation of momentum-driven winds is rather
crude. For example, the wind reaches its maximum 
velocity when it leaves the ISM rather than in the outer halo as
expected for radiatively driven winds. Moreover, the parametrization
leaves a lot of freedom, even 
more than for SN feedback, partly because the total
amount of energy is no longer limited\footnote{Depending on redshift, 
\emph{WVCIRC} injects more energy in the winds per unit stellar mass
formed than model \emph{REF} for halo masses that exceed $10^{11} -
10^{12}\,\Msun$.}. More to the point, Haas et al.\ (in preparation)
show that the models inject much more momentum than is actually
available in the form of star light.
It would therefore be dangerous to use comparisons
between observations and models such as these to discriminate between
outflows driven by SNe and radiation pressure. Moreover, we note that
it is not clear that observations 
of high-mass galaxies should agree with the ``momentum-driven''
models, given that AGN feedback is thought to be crucial for such
objects
\citep[e.g.][]{Croton2006SA,Bower2006SA},
but was not included. Conversely, if simulations such as the ones
presented here were to disagree with observations, it would not
necessarily mean that winds are not radiatively driven.

It is interesting that the scalings of the ``momentum-driven''
prescription, i.e.\ a wind velocity that increases 
and a mass loading that decreases with galaxy mass, agree
qualitatively with the results obtained for high-resolution
simulations of energy-driven SN feedback
\citep{DallaVecchia2008Winds,DallaVecchia2009Winds}. Even when
hydrodynamical interactions are not temporarily ignored, the mass
loading will be underestimated for galaxies that are poorly
resolved \citep{DallaVecchia2008Winds}. It may therefore be that the
``momentum-driven'' wind scaling partly compensates for some of the
resolution effects that plague cosmological simulations, particularly
at high redshift.

\subsection{AGN Feedback}
\label{sec:agn}

\begin{figure}
\resizebox{\colwidth}{!}{\includegraphics{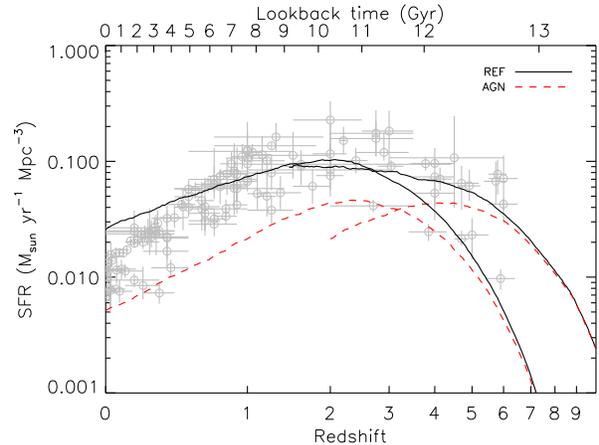}}
\caption{As Fig.~\protect\ref{fig:cosmology},
but comparing the SFHs for models with and without supermassive
BHs. Feedback from AGN strongly suppresses the SFR in massive
galaxies, which becomes more important at late times. When AGN are
included the SFR appears to be too low compared with observations, but that
could be changed by decreasing the fraction of the SN energy that is
injected, which was in this case chosen to roughly match the peak of
the SFH in the absence of AGN feedback.}  
\label{fig:agn}
\end{figure}

The centers of galaxies are thought to harbor supermassive BHs. Matter
accreting onto these BHs emits large amounts of high 
energy radiation. Even if only a small fraction of this energy gets
coupled to the ISM, it could have a dramatic effect on the
galaxies. Moreover, the magnetic fields carried by the accreting
matter could lead to the formation of jets which can displace and
heat gas in and around galaxies. Feedback from AGN has for example been invoked to
explain the low SFRs of high-mass galaxies
\citep[e.g.][]{DiMatteo2005AGN,Croton2006SA,Bower2006SA,Booth2009AGN}
and the suppression of cooling flows in clusters of galaxies
\citep[e.g.][]{Churazov2001AGN,DallaVecchia2004AGN,Ruszkowski2004AGN,Cattaneo2007AGN}.   

To investigate the effect of AGN feedback, we have run a series of
simulations employing the subgrid prescription for the growth of BHs
and feedback from AGN described in \citet{Booth2009AGN} which is a
substantially modified version of the model of
\citet{Springel2005AGN}. Below we will briefly summarise the main
features of this model, but we refer the reader to
\citet{Booth2009AGN} for further details and tests.  

Seed BHs of mass $m_{\rm seed}$ are placed into every
dark matter halo whose mass exceeds $m_{\rm halo,min}$. Our fiducial
AGN model uses $m_{\rm seed} = 9\times 10^4\,\Msun$, which corresponds
to $10^{-3}\,m_{\rm g}$ ($6.4\times 10^{-2}\,m_{\rm g}$) in the
100~$\hMpc$ (25~$\hMpc$) box. For the minimum dark halo mass we use
$m_{\rm halo,min}=4\times 10^{10}\,\Msun$, which corresponds to $10^2$ ($6.4\times
10^3$) dark matter particles in the 100~$\hMpc$ (25~$\hMpc$)
box. Haloes are identified by running a friends-of-friends group
finder  
on-the-fly as described in section~\ref{sec:wmom}. BHs can grow via  
Eddington-limited accretion of the surrounding gas and through mergers  
with other BHs. 
  
\citet{Booth2009AGN} show that any simulation that
resolves the Jeans scales will also resolve the Bondi-Hoyle accretion
radius for BHs whose mass exceeds the simulation's mass
resolution. At densities below the SF threshold $n_{\rm H}^\ast$ 
($10^{-1}\,\cm^{-3}$ in our model), the gas is kept warm by the
ionizing background \citep{Schaye2004SF} and we marginally resolve the
Jeans scales in our highest resolution runs (see
Section~\ref{sec:convergence}). For higher densities, however, a cold
phase is expected to be present and naive application of the
Bondi-Hoyle formula would lead us to strongly underestimate the
accretion rate. We therefore assume that the accretion rate is given
by the minimum of the Eddington rate and 
\begin{equation}
 \dot{m}_{\rm accr} = \alpha
\frac{4\pi G^2 m_{\rm BH}^2 \rho}{(c_{\rm s}^2+v^2)^{3/2}},
 \label{eq:bhl}
\end{equation}
where $m_{\rm BH}$ is the mass of the BH, $c_{s}$ and $\rho$ are the
sound speed and density of the local medium, $v$ is the velocity of
the BH relative to the ambient medium, and $\alpha$ is a dimensionless
efficiency parameter given by\footnote{Note
that \citet{Springel2005AGN} used a fixed value
$\alpha=100$. Consequently, massive
BHs need to suppress the ambient gas density to values far below
$n_{\rm H}^\ast$ in order to reduce the accretion rate to
sub-Eddington values \citep[see][]{Booth2009AGN}.}
\begin{equation}
 \alpha=\left\{ \begin{array}{cc} 1 & {\rm if~} n_{\rm H}<n_{\rm
       H}^* \\ \Big(\frac{n_{\rm H}}{n_{\rm H}^*}\Big)^\beta &
   \textrm{otherwise.}\end{array}\right.
 \label{eq:beta}
\end{equation}
Observe that for $\alpha=1$ (i.e.\ $n_{\rm H}<n_{\rm H}^\ast$ or
$\beta=0$), equation~(\ref{eq:bhl}) reduces to the
Bondi-Hoyle-Lyttleton rate \citep{Bondi1944Accr,Hoyle1939Accr}. Our
fiducial AGN runs use $\beta=2$, which results in efficient BH growth
in haloes with stellar masses $\ga 10^{10.5}\,\Msun$ in the
\emph{L100} runs. 

The amount of
accreted mass is related to the rate of growth of the BH by $\dot{m}_{\rm BH} =
\dot{m}_{\rm accr} (1-\epsilon_{\rm r})$, where $\epsilon_{\rm r}$ is
the radiative efficiency of a BH, 
which we always assume to be 10\%, the mean value for the radiatively
efficient \citet{Shakura1973RadEff} accretion onto a Schwarzschild BH.

We assume that a fraction $\epsilon_{\rm f}$ of the radiated energy
couples to the ISM. The amount of energy returned by
a BH to its surrounding medium is thus given by
\begin{equation}
\label{eq:epsilon}
 \dot{E}_{\rm feed}=\epsilon_{\rm f} \epsilon_{\rm r} \dot{m}_{\rm accr}
 c^2,
\end{equation}
where $c$ is the speed of light. We set $\epsilon_{\rm f}=0.15$ in
order to match the observed cosmic mass density in BHs as well as the
relation between BH and galaxy mass, both at redshift
zero. BH particles store feedback energy until it suffices to heat $n_{\rm
  heat}$ of their neighbours by $\Delta T_{\rm 
    min}$. The
two parameters $\Delta T_{\rm min}$ and $n_{\rm heat}$ are chosen such
that AGN heated gas obtains a long cooling time and so that the time
taken to perform a feedback event is shorter than the Salpeter
time for Eddington-limited accretion. The parameter choices $\Delta T_{\rm
  min}=10^8\,\K$ and  $n_{\rm heat}=1$ are found to provide a good
balance between these two constraints.

Fig.~\ref{fig:agn} shows that the addition of AGN
feedback strongly  
suppresses the SFR. The difference with the
reference model increases 
with time, which implies that AGN feedback is more important for
higher mass galaxies. The drop in the SFR below $z=2$ is much closer
to the observed slope when AGN feedback is included, but the overall
amplitude of the SFR is probably too low. It is, however, important to
note that the observed SFRs are subject to large systematic
uncertainties. Indeed, McCarthy et al.\ (in preparation) find that the
predicted stellar masses are in fact in good agreement with
observations of groups of galaxies. 

Even if the simulation with AGN feedback really did form too few
stars, it would not be a 
concern here. As discussed in section~\ref{sec:Ref_sn}, the fraction of the
SN energy that is injected was chosen to roughly match the peak in the
observed SFH. It is therefore unavoidable that including an extra form
of efficient feedback, without making any other changes, reduces the
SFR to values that are lower than observed. It would have
been possible to adjust the parameters of the prescription for SN
feedback to obtain a better match to the observed SFH, but this is not
the objective here. 

\citet{Booth2009AGN} have carried out an extensive parameter study of
the AGN model at the resolution of our 100~$\hMpc$ box, comparing
the cosmic SFH as well as other observables. They found that the
results are sensitive to the accretion model, i.e.\ the value of
$\beta$, which sets the halo mass above which the BHs can grow onto
the scaling relations. Remarkably, they found that the SFH is nearly
completely independent of the feedback efficiency, but, in agreement
with \citet{DiMatteo2005AGN}, the BH masses
are inversely proportional to $\epsilon_{\rm f}$. They explained this
in terms of self-regulation: the BHs grow 
until they have injected sufficient energy to balance
the infall driven by gas accretion onto haloes and radiative
cooling \cite[see
  also][]{DiMatteo2005AGN,Robertson2006msigma,Booth2009BHmasses}. If half as
much energy is injected per unit  
accreted mass, then the BHs need to grow twice as massive in order to
inject the same amount of energy. Because the factor by which SF is
suppressed depends on the amount of energy that is injected by the
BHs, the SFH is insensitive to variations in the assumed efficiency of
AGN feedback.

\section{Summary and discussion}
\label{sec:summary}

The cosmic star formation history (SFH) is perhaps the most
fundamental observable in 
astrophysical cosmology. It is difficult to model, because of the
large range of galaxy masses that contribute and because of the many
feedback processes that may be important. Cosmological hydrodynamical
simulations need to resort to subgrid prescriptions for the physics
that remains unresolved, which makes them resemble semi-analytic
models in some respects. It is, however, much more difficult to
explore parameter space using fully numerical simulations because of
the high computational cost.  

Here, we have introduced the OverWhelmingly Large Simulations (OWLS)
project, which consists of more than 50 large, cosmological,
hydrodynamical simulations. The simulations were all run with a
modified version of the SPH code \textsc{gadget3} \citep[last
  described in][]{Springel2005Gadget2}, using new modules for
radiative cooling \citep{Wiersma2009Cooling}, SF \citep{Schaye2008SF},
chemodynamics \citep{Wiersma2009Chemo}, kinetic
\citep{DallaVecchia2008Winds} or thermal \citep{DallaVecchia2009Winds}
SN feedback, and accretion onto and feedback from supermassive
BHs \citep{Springel2005AGN,Booth2009AGN}. With $2\times 512^3$
particles, the OWLS runs are among the largest dissipative
simulations ever performed. The simulations
are repeated many times, each time changing a single
aspect of the input 
physics or a single numerical parameter with respect to the reference
model described in section~\ref{sec:reference}. We stress that this
model merely functions as a reference point for our systematic
exploration of parameter space and should therefore not be regarded as
our ``best'' model.  

Generically, we find that SF is limited
by the build-up of dark matter haloes at high redshift, reaches a broad
maximum at intermediate redshift, then decreases as it is quenched by
lower cooling rates in hotter and lower density gas, gas exhaustion,
and self-regulated feedback from stars and black holes, in broad
agreement with previous work \citep[e.g.][]{White1991GF,Hernquist2003SFH}.
While we have limited ourselves
to comparisons of the cosmic SFHs, future papers will investigate the
behavior of the models in other contexts. 

The models are run
down to redshift $z=2$ in boxes of 25~comoving $\hMpc$ on a side and/or
down to $z=0$ in boxes of 100~$\hMpc$. The mass (spatial) resolution
is a factor 64 (4) better in the smaller box. 
Detailed convergence tests showed that a $25~\hMpc$ box is 
sufficiently large to model the cosmic SFH down to
$z=0$. For the 25~$\hMpc$
(100~$\hMpc$) simulation of the reference model, the SFH is very close to
converged for $z<7$ ($z<3$). Interestingly, we found that while the
star formation rate (SFR) typically increases if the resolution is
improved, the situation 
reverses at low redshift before convergence has been attained. This
calls into question the strategy to combine  different box sizes to
obtain a converged SFH all the way from high redshift to $z=0$.

The cosmic SFR density can be decomposed into a halo mass function and
the (distribution of the) SFR as a function of halo mass. The
baryonic physics mainly affects the latter function, which we shall
explore in more detail in Haas et al.\ (in preparation), while the mass function is
determined mostly by the assumed cosmology. We showed that the SFH is
sensitive to even relatively
small changes in the cosmological parameters, such as the difference
between the values inferred from the WMAP 3-year data used
here and the earlier values assumed in the Millennium
simulation. Clearly, the parameters of semi-analytic models build onto
the Millennium 
simulation will have to be modified if they are used on a simulation
assuming the current concordance cosmology. Comparisons to
observations of rare objects and the  
high-redshift Universe will be particularly strongly affected
due to the relatively large difference in the value of $\sigma_8$. 

Our systematic tests of the subgrid physics revealed that SF in
intermediate mass galaxies is highly self-regulated by feedback from
massive stars. This explains our remarkable
finding that the predicted SFH is nearly completely independent of the
treatment of the unresolved ISM, including the assumed SF law. If the
SF efficiency is increased, then the galaxies 
simply reduce their gas 
fractions so as to keep the SFR, and thus the rate at which stars
inject energy into the ISM, constant. Similarly, if the efficiency of SN
feedback is changed by injecting a different fraction of the SN
energy, then, to first order, the galaxies simply adjust their SFRs so
as to keep the rate of energy injection constant. The critical rate of
energy injection that results from self-regulation is presumably the
rate required to balance gas infall resulting from accretion onto
haloes and radiative cooling and will therefore depend on the
halo mass and redshift. 

Note, however, that our findings apply to SFRs that have effectively
been averaged over entire galaxies and over very long time scales.
Self-regulation may also occur on smaller length and time
scales. Indeed, we implicitly assume this to be the case through our
use of empirical SF laws that have been averaged over spatial scales
that are large compared to individual star-forming regions. Processes
other than SN feedback may be important for the self-regulation that
occurs on these smaller scales and perhaps even on large scales if SN
feedback is inefficient
\citep[e.g.][]{Robertson2008SFlaw,Gnedin2009H2form}.  

For massive galaxies feedback from massive stars becomes inefficient
and it is the self-regulated growth of
supermassive BHs that ensures that a fixed amount of energy is
injected into the ISM
\citep[e.g.][]{DiMatteo2005AGN,Booth2009AGN,Booth2009BHmasses}. As we
showed in \citet{Booth2009AGN}, 
the SFH is therefore independent of the assumed efficiency of AGN
feedback. If the BHs inject twice as
much energy per unit accreted mass, then they grow only half as
massive so that they inject the same amount of energy and because they
inject the same amount of energy, the SFR is suppressed by the same
factor. 

At very high redshift the SFH is not yet controlled by
self-regulation and will thus be limited by the gas consumption time
scale implied by the assumed SF law.
In our simulations the dependence on the SF law at high redshift may,
however, mostly result from our limited resolution.
Stars can only form in haloes that exceed the resolution
limit and the formation of the first generation of stars within a halo
is not hampered by winds. Because the minimum halo mass is set by the 
resolution, so is the SFH until it is dominated by haloes that are
significantly more massive. 

Radiative feedback from non-local SF is, however, very important
at high redshift. In particular, reheating
associated with hydrogen reionization quenches SF in haloes with
virial temperatures $\la 10^4\,\K$ and thus strongly affects the SFH
when it would otherwise be dominated by such low-mass
haloes. Our 
simulations underestimate this effect due to their limited
resolution and because we assume that a photo-dissociating background
is present at all redshifts. Contrary to hydrogen reionization, the
milder reheating associated with helium reionization does not have a
significant impact on the SFH because it is limited to the low-density
IGM. 

While cosmology and self-regulation via outflows are the principal
ingredients controlling the SFH, there are other processes that are
important. Metal-line
cooling becomes very important at late times, as the metal content of
the gas builds up and less gas falls in cold. The inclusion of cooling
by heavy elements 
has two important effects. First, it allows more virialized gas to
cool, an effect that has been widely discussed in the 
literature. However, we find that the
dominant effect of metal-line cooling may be that it reduces the
efficiency of galactic winds, as it increases thermal losses in the
gas that has been shock-heated by the winds. The inclusion of metal-line cooling
makes it much more difficult to reproduce the steep decline in the
observed SFH below $z=2$.

Another process that becomes important at late times, is mass
loss by intermediate mass stars. On time scales of hundreds of
millions to billions of years, winds from AGB stars boost the SFR by
providing fresh
fuel for SF and by releasing metals that were locked up in stars into
the ISM. As was the case for metal-line cooling, including mass loss
by AGB stars make it more difficult to match the sharp drop in the
cosmic SFR that is observed at low redshift. 
The shape of the time delay function for SNe of type Ia, on
the other hand, turned out to be unimportant, presumably because its
impact is limited to a shift in the timing of the release of a
fraction of the iron.  

Without AGN feedback, it is challenging to match the steep decline in
the cosmic SFR below $z=2$. It is, however, not yet
totally clear that feedback from 
massive stars cannot solve the problem. If injected in kinetic form,
SN feedback becomes 
inefficient once the wind velocity falls below some critical value
that increases with galaxy mass and thus with the pressure in the ISM
\citep{DallaVecchia2008Winds}. By increasing the input wind velocity with
galaxy mass, or with some correlated local property (e.g.\ gas
density, pressure, or velocity dispersion), while decreasing the input mass
loading so as to keep the wind energy constant, we can keep SN feedback
efficient in higher mass galaxies. 

Using a sufficiently high constant
wind velocity also results in efficient suppression of SF in
relatively massive galaxies, but not
for poorly resolved low-mass galaxies. If a galaxy is poorly resolved,
then there is no disc from which wind particles can drag gas along,
limiting the effective mass loading factor to the input value 
\citep{DallaVecchia2008Winds}. Decreasing the input wind velocity (and
thus increasing the input
mass loading) with decreasing galaxy mass counteracts the resulting
underestimate of the mass loading factor.
If we allow ourselves to vary the
parameters of the kinetic feedback with local properties, then we
can even ``design'' SFHs, but this freedom comes at the expense of
introducing additional free parameters and an increased sensitivity to
the numerical resolution.   

Recently, the possibility that galactic outflows are driven by
radiation pressure on dust grains has generated considerable interest
\citep{Murray2005MomWinds}. For such winds the mass loading is
expected to be inversely proportional to the wind velocity, which, in
the outer halo, is
expected to be similar to the galaxy velocity dispersion.
\citet{Oppenheimer2006Wpot,Oppenheimer2008Wmom}
 implemented such ``momentum-driven''
winds in a simplified fashion by kicking particles out of the ISM with
several times the velocity that the wind is expected to reach in the
outer halo, and by tuning the 
normalization of the initial wind mass loading to match the observed
SFR. We ran several simulations that employed such methods and found
that the feedback is more efficient than our
standard SN feedback, particularly at low redshift when relatively massive
galaxies dominate the SFR. This is partly because the winds
are allowed to carry more energy than is available from SNe, but it
may be mostly due to the fact that the ``momentum-driven'' scalings
happen to overcome 
some of the numerical effects discussed above. Given the simplified
nature and the limitations of the kinetic feedback models, it would be
dangerous to use them to discriminate between winds driven by SNe and
radiation pressure. Finally, we note that Haas et al.\ (in
preparation) demonstrate that the amount of momentum that is injected
in the models of \citet{Oppenheimer2006Wpot,Oppenheimer2008Wmom}, and
thus also in our versions of these models, exceeds the amount of momentum
that is actually available in the form of star light by up to an order
of magnitude.

Besides the values of the wind parameters, the implementation of
kinetic SN feedback is also important. For example, most studies in
the literature employing \textsc{gadget} temporarily decouple wind
particles from the hydrodynamics, allowing them to freely escape the
ISM without blowing bubbles or generating turbulence. We found that
this decoupling (which we included in one of our runs), reduces the
efficiency of SN feedback at high redshift, because of the inability
of the winds to drag gas along, but strongly increases it at low
redshift when hydrodynamic drag would otherwise quench the winds in
massive galaxies. 

Outflows driven by feedback from SF, be it SN
or radiation pressure, are very
important. It is therefore unfortunate that the predictions for the
SFH of simulations
that implement outflows in the form of kinetic feedback, i.e.\ by
kicking particles, are not robust at the currently attainable
resolution. The predicted SFH is sensitive to the values 
of poorly constrained wind parameters, even for a constant wind
energy, and to the details of the implementation. Note that the same
may well be true for other types of subgrid prescriptions for feedback
from SF. Clearly, it is also crucial to vary the parameters of such
models. On the other hand, we will show in future papers that many
observables are much less sensitive to the implementation of
feedback from SF than is the case for the cosmic SFH.

We investigated implementing SN feedback in thermal form, using the
method of \citet{DallaVecchia2009Winds} to overcome the 
overcooling problem that is commonly encountered when thermal SN
feedback is used. Encouragingly, we found thermal feedback to be
efficient and the predictions to be less sensitive to the values of the free
parameters than is the case for kinetic feedback. Relative to kinetic
feedback, thermal feedback is 
particularly efficient at high redshift. Because heated particles push
all their neighbours, the winds remain highly mass loaded even in
poorly resolved galaxies. 

Another large uncertainty is the stellar IMF. The consequences of a
change in the assumed IMF are difficult to predict because the IMF affects
many things. The IMF determines
the effective nucleosynthetic yields, the fraction of the stellar mass
that is 
recycled by intermediate mass stars, and the amount of SN energy (and
radiation) produced by massive stars. Moreover, a change in the IMF
also implies a change in the observed SF law and history, as SFRs are
inferred from observations of the light 
emitted by massive stars. These latter two effects are often
ignored. Interestingly, we found that an IMF that becomes top-heavy at
high gas pressures would improve agreement with the observations
because it preferentially suppresses SF in massive galaxies. 

How do we proceed from here? Based on our findings, a better
understanding of the mechanisms responsible for the generation of
galactic winds, as well as more robust numerical implementations, are
crucial. Paradoxically, a better understanding of SF may not
directly improve predictions for the cosmic SFH if, as is the case for
the models studied here, galaxies
regulate their SFRs by adjusting their gas fractions via large-scale
outflows. At very high redshift 
photo-heating is key and simulations including radiative transfer
would constitute a clear improvement. At lower redshifts it is very important
to include mass loss from AGB stars and metal-line cooling. 
Better treatments of metal 
mixing and the inclusion of non-equilibrium 
cooling would certainly be helpful here. The steep drop in the
observed SFR below 
$z=2$ is difficult to reproduce without AGN feedback. Fortunately, the
self-regulatory nature of the growth of supermassive BHs makes
predictions for the SFH insensitive to the assumed efficiency of AGN feedback.
Finally, as always, higher resolution would be very helpful as it
would, for example, enable us to probe further down the halo mass
function and to include more realistic treatments of galactic winds.

\section*{Acknowledgments}

It is a great pleasure to thank Volker Springel without whose generous
help this project would not have been possible. We are
also grateful to Anthony Aguirre, Richard Battye, Stefano Borgani, 
Richard Bower, Kathy Cooksey, 
Rob Crain, Scott Kay, Julio Navarro, Benjamin 
Oppenheimer, Andreas Pawlik, Xavier Prochaska, Olivera Rakic, Philipp
Richter, Laura Sales, Debora 
Sijacki, Thorsten Tepper-Garc\'ia, Luca Tornatore, and all the members
of the Virgo 
consortium for discussions and help. We would also like to thank the
referee, Brant Robertson, for a constructive report.
The simulations presented here were run on Stella, the LOFAR
BlueGene/L system in Groningen, on the Cosmology Machine at the
Institute for Computational Cosmology in Durham as part of the Virgo
Consortium research programme, and on Darwin in Cambridge. This work
was sponsored by National Computing Facilities Foundation (NCF) for
the use of supercomputer facilities, with financial support from the
Netherlands Organization for Scientific Research (NWO). 
This work was supported by Marie Curie Excellence Grant
MEXT-CT-2004-014112 and by an NWO VIDI grant. IGM acknowledges support
from a Kavli Institute Fellowship at the University of Cambridge. 

\bibliographystyle{mn2e} 
\bibliography{ms}

\label{lastpage}

\end{document}